\documentclass{article}

\usepackage{microtype}
\usepackage{graphicx}
\usepackage{booktabs} % for professional tables

\usepackage{hyperref}
\usepackage{textcomp}

\usepackage[accepted]{icml2023}

\usepackage{amsmath}
\usepackage{amssymb}
\usepackage{mathtools}
\usepackage{amsthm}

\usepackage[capitalize,noabbrev]{cleveref}

\usepackage{multirow}
\usepackage{algorithmic}
\usepackage{subcaption}
\newsavebox{\largestimage}
\usepackage[nodisplayskipstretch]{setspace}

\theoremstyle{plain}
\newtheorem{theorem}{Theorem}[section]

\newtheorem{lemma}[theorem]{Lemma}

\theoremstyle{definition}

\theoremstyle{remark}
\newtheorem{remark}[theorem]{Remark}

\usepackage[textsize=tiny]{todonotes}

\newcommand{\vect}[1]{\mathbf{vec}\left(#1\right)}

\DeclareMathOperator*{\argmin}{arg\,min}
\newcommand{\angstrom}{\mathrm{\normalfont\AA}}
\newcommand\numberthis{\addtocounter{equation}{1}\tag{\theequation}}

\icmltitlerunning{Tensor Gaussian Process with Contraction for Multi-Channel Imaging Analysis}

\begin{document}

\twocolumn[
\icmltitle{Tensor Gaussian Process with Contraction for Multi-Channel Imaging Analysis}

% It is OKAY to include author information, even for blind
% submissions: the style file will automatically remove it for you
% unless you've provided the [accepted] option to the icml2022
% package.

% List of affiliations: The first argument should be a (short)
% identifier you will use later to specify author affiliations
% Academic affiliations should list Department, University, City, Region, Country
% Industry affiliations should list Company, City, Region, Country

% You can specify symbols, otherwise they are numbered in order.
% Ideally, you should not use this facility. Affiliations will be numbered
% in order of appearance and this is the preferred way.
\icmlsetsymbol{equal}{*}

\begin{icmlauthorlist}
\icmlauthor{Hu Sun}{stats}
\icmlauthor{Ward Manchester}{CLASP}
\icmlauthor{Meng Jin}{LM}
\icmlauthor{Yang Liu}{SDO}
\icmlauthor{Yang Chen}{stats}
\end{icmlauthorlist}

\icmlaffiliation{stats}{Department of Statistics, University of Michigan, Ann Arbor}
\icmlaffiliation{CLASP}{Department of Climate and Space Sciences and Engineering, University of Michigan, Ann Arbor}
\icmlaffiliation{LM}{Solar \& Astrophysics Lab, Lockheed Martin}
\icmlaffiliation{SDO}{W.W. Hansen Experimental Physics Laboratory, Stanford University}

% \icmlcorrespondingauthor{Hu Sun}{husun@umich.edu}
\icmlcorrespondingauthor{Yang Chen}{ychenang@umich.edu}

% You may provide any keywords that you
% find helpful for describing your paper; these are used to populate
% the "keywords" metadata in the PDF but will not be shown in the document
\icmlkeywords{Tensor Regression, Tensor Contraction, Gaussian Process}

\vskip 0.3in
]

% this must go after the closing bracket ] following \twocolumn[ ...

% This command actually creates the footnote in the first column
% listing the affiliations and the copyright notice.
% The command takes one argument, which is text to display at the start of the footnote.
% The \icmlEqualContribution command is standard text for equal contribution.
% Remove it (just {}) if you do not need this facility.

%\printAffiliationsAndNotice{}  % leave blank if no need to mention equal contribution
\printAffiliationsAndNotice{} % otherwise use the standard text.

\begin{abstract}
Multi-channel imaging data is a prevalent data format in scientific fields such as astronomy and biology. The structured information and the high dimensionality of these 3-D tensor data makes the analysis an intriguing but challenging topic for statisticians and practitioners. The low-rank scalar-on-tensor regression model, in particular, has received widespread attention and has been re-formulated as a tensor Gaussian Process (Tensor-GP) model with multi-linear kernel in \citet{yu2018tensor}. In this paper, we extend the Tensor-GP model by introducing an integrative dimensionality reduction technique, called \textit{tensor contraction}, with a Tensor-GP for a scalar-on-tensor regression task with multi-channel imaging data. This is motivated by the solar flare forecasting problem with high dimensional multi-channel imaging data. We first estimate a latent, reduced-size tensor for each data tensor and then apply a multi-linear Tensor-GP on the latent tensor data for prediction. We introduce an anisotropic total-variation regularization when conducting the tensor contraction to obtain a sparse and smooth latent tensor. We then propose an alternating proximal gradient descent algorithm for estimation. We validate our approach via extensive simulation studies and applying it to the solar flare forecasting problem.
\end{abstract}

\section{Introduction}
Regression models that deal with scalar labels and tensor covariates, i.e. scalar-on-tensor regression, have received widespread attention over the past decade \cite{hung2013matrix,zhou2013tensor,zhou2014regularized,kang2018scalar,li2018tucker,papadogeorgou2021soft}. Given $m$-mode tensor covariate $\mathcal{X}\in\mathbb{R}^{I_{1}\times I_{2} \times \dots \times I_{m}}$ and scalar label $y\in \mathbb{R}$, the existing literature approaches the regression problem mainly via:
\begin{equation}\label{eq:tensor-regression}
    \mathbb{E}[y|\mathcal{X}] = \alpha + \left<\mathcal{W}, \mathcal{X}\right>, 
\end{equation}
where $\alpha$ is the intercept, $\mathcal{W}$ is the regression coefficient tensor that matches the shape of $\mathcal{X}$ and $\left<\cdot,\cdot\right>$ denotes tensor inner product following \citet{kolda2009tensor}. This formulation can be readily adopted under the framework of generalized linear model \cite{zhou2013tensor} while simultaneously preserving the tensor structure of $\mathcal{X}$. Typically, tensor data is of ultra-high dimensions and thus $\mathcal{W}$ is also of high dimensionality. Various constraints have been introduced on $\mathcal{W}$, such as tensor norm regularization \cite{guo2011tensor,zhou2014regularized} and tensor rank constraints \cite{papadogeorgou2021soft,hao2021sparse}. These constraints induce a sparse and low-rank structure over $\mathcal{W}$, making inferences of the high-order correlation between the scalar label and the tensor covariates tractable and interpretable.

Gaussian Process (GP) \cite{williams2006gaussian} is an alternative approach to modeling complex correlation structures, and has been applied to tensor regression problems \cite{kang2018scalar}, where a GP prior is imposed on $\mathcal{W}$. In \citet{yu2018tensor}, it is established that the tensor regression model \eqref{eq:tensor-regression}, together with a low-rank constraint on $\mathcal{W}$, leads to the same estimator $\widehat{\mathcal{W}}$ as the tensor Gaussian Process (\textbf{Tensor-GP}) coupled with a multi-linear kernel on the prior of $\mathcal{W}$. A multi-linear kernel function $k(\cdot,\cdot)$ for $m$-mode tensors $\mathcal{X} \in \mathbb{R}^{I_{1}\times I_{2}\times \dots \times I_{m}}$ can be defined in a Kronecker product form as:
\begin{equation*}
    k(\mathcal{X}_{i}, \mathcal{X}_{j}) = \vect{\mathcal{X}_{i}}^{\top} \left(\otimes_{m^{'}=1}^{m} \mathbf{K}_{m+1-m^{'}}\right) \vect{\mathcal{X}_{j}},
\end{equation*}
where $\vect{\cdot}$ is the vectorization operator and $\otimes$ denotes the matrix Kronecker product and $\mathbf{K}_{1}, \dots, \mathbf{K}_{m}$ capture the mode-specific covariance structure of the regression coefficient tensor $\mathcal{W}$ and are assumed to be low-rank. Interpreting this GP regression model can be hard since one needs to inspect the multi-linear kernel which deals with the tensor data at its original dimensionality $d = \prod_{m^{'}=1}^{m} I_{m^{'}}$. 

The capability of the multi-linear Tensor-GP to provide uncertainty quantification on the prediction makes it an attractive alternative to its counterpart in \eqref{eq:tensor-regression}, but a sufficient dimension reduction on the tensor data is needed to make it more interpretable for scientific applications. In a different thread of literature, in \citet{kossaifi2020tensor}, a tensor contraction operation is introduced before estimating the tensor regression model under the neural network settings. Instead of compressing the information of tensor data into a vector,
% like most deep neural networks (e.g. Variational Auto-encoder \cite{kingma2013auto}), 
the tensor data is contracted into a smaller \textit{core} tensor with the same number of modes. Such a dimension reduction technique preserves the tensor structure of the data, making tensor regression or Tensor-GP directly applicable.

In this paper, we propose a novel framework by combining the merits of tensor contraction and Tensor-GP for the scalar-on-tensor regression task. Our framework consists of two major blocks. Firstly, we introduce tensor contraction to transform the tensor data $\mathcal{X}$ to a feature tensor $\mathcal{Z}$ with much lower dimensionality. Secondly, we apply the multi-linear Tensor-GP to the reduced-sized tensor $\mathcal{Z}$ for regression. We build our model around a special type of tensor, i.e. the multi-channel imaging tensor, motivated by an application to astrophysical imaging analysis. But our model can be easily extended to a general tensor setup. We summarize our contributions as follows:

\begin{itemize}
    \item We integrate tensor dimension reduction with Tensor-GP in a unified framework called \textbf{Tensor-GPST}, allowing for learning a low-dimensional tensor representation in a supervised learning context.
    \item We propose to use the anisotropic total variation regularization \cite{wang2017generalized} in the tensor contraction step for a sparse and spatially smooth tensor dimension reduction. We estimate the parameters of Tensor-GPST jointly under a penalized marginal likelihood approach coupled with the proximal gradient method \cite{parikh2014proximal} with convergence guarantee.
\end{itemize}

% Tensor data such as the brain-imaging data and the spatial-temporal processes data typically have ultra-high dimensionality (e.g. $\ge 100k$) in contract with the very limited sample size (e.g. $\le 1k$), leading statisticians to introduce additional constraints to reduce the degrees of freedom of $\mathcal{W}$.

\section{Tensor Gaussian Process with Spatial Transformation (Tensor-GPST)}
In this section, we will first introduce our method, called Tensor-GPST, for the scalar-on-tensor regression task and then discuss the algorithm in section \ref{subsec:algo} for estimating its parameters and conclude by discussing the theoretical guarantee of the algorithm convergence in section \ref{subsec:theory}.

Throughout the paper, we use calligraphic letters (e.g. $\mathcal{X},\mathcal{Z}$) for tensors with at least three modes, boldface uppercase letters (e.g. $\mathbf{A}, \mathbf{B}$) for matrices, boldface lowercase letters (e.g. $\mathbf{w}$, $\mathbf{y}$) for vectors and plain letters (e.g. $\lambda, s$) for scalars. For an $m$-mode tensor $\mathcal{X}$ of size $I_{1}\times I_{2}\times \dots \times I_{m}$, its $k$-mode product with matrix $\mathbf{U}\in\mathbb{R}^{J\times I_{k}}$, denoted as $\mathcal{X}\times_{k} \mathbf{U}$, is an $m$-mode tensor of size $I_{1}\times \dots I_{k-1}\times J \times I_{k+1}\times \dots \times I_{m}$, where:
\begin{equation*}
    \left(\mathcal{X} \times_{k} \mathbf{U}\right)_{i_{1},\ldots,j,\ldots,i_{m}} = \sum_{i_{k}=1}^{I_{k}} \mathcal{X}_{i_{1},\ldots,i_{k},\ldots,i_{m}} \mathbf{U}_{ji_{k}}.
\end{equation*}
We use $\left<\mathcal{X},\mathcal{Y}\right>$ to denote tensor inner product and $\|\mathcal{X}\|_{\mathrm{F}} = \sqrt{\left<\mathcal{X},\mathcal{X}\right>}$ to denote tensor Frobenius norm. We refer the readers to \citet{kolda2009tensor} for a thorough introduction to tensor algebra.

\subsection{Method}\label{subsec:Method}
We consider a multi-channel imaging dataset $\{\mathcal{X}_{i}, y_{i}\}_{i=1}^{N}$, where $\mathcal{X}_{i} \in \mathbb{R}^{H\times W\times C}$ with $H, W, C$ as the height, width and number of channels, respectively; and $y_{i}\in \mathbb{R}$. We use $\mathbf{X}_{i}^{(c)} \in \mathbb{R}^{H\times W}, c\in[C]$ to denote the $c^{\rm{th}}$ channel of $\mathcal{X}_i$. 
% We use $\mathbf{X} \in \mathbb{R}^{N\times  H\times W\times C}$ and $\mathbf{y} \in \mathbb{R}^{N}$ to denote the concatenated imaging data and regression labels. 
Gaussian process regression (GPR) \cite{williams2006gaussian} specifies the prior for $y_i$ as:
\begin{equation}\label{eq:Baseline-GP}
    y_i = f(\mathcal{X}_{i}) + \epsilon_{i}, \quad f(\cdot) \sim \mbox{GP}\left(m(\cdot), k(\cdot,\cdot)\right),
\end{equation}
with $\epsilon_{i} \sim \mathcal{N}\left(0, \sigma^{2}\right)$ being the idiosyncratic noise. The GP prior characterizes the unknown function $f(\cdot)$ evaluated at all data points as a multivariate Gaussian distribution, with a mean function $m(\cdot)$ and a covariance kernel function $k(\cdot,\cdot)$. Typically, $m(\cdot)$ is assumed to be zero and $k(\cdot,\cdot)$ fully specifies the behavior of the GP prior.

Given the high dimensionality of $\mathcal{X}_{i}$, it would be difficult to directly estimate and interpret the tensor kernel $k(\cdot,\cdot)$. Here we consider adding one extra step called \textit{tensor contraction}, which compresses the information of $\mathcal{X}_{i} \in \mathbb{R}^{H\times W\times C}$ into a reduced-sized tensor $\mathcal{Z}_{i} \in \mathbb{R}^{h\times w\times C}$, with $h<H, w<W$, via:
\begin{equation}\label{eq:tensor-contraction}
    \mathcal{Z}_{i} = g(\mathcal{X}_{i}) = \mathcal{X}_{i} \times_{1} \mathbf{A} \times_{2} \mathbf{B} \times_{3} \mathbf{I}_{C},
\end{equation}
with $\mathbf{A} \in \mathbb{R}^{h\times H}, \mathbf{B} \in \mathbb{R}^{w\times W}$. 
% For the $(i_1,\ldots,i_{k-1},j,i_{k+1},\ldots,i_{K})$ entry of $\mathcal{X}\times_{m} U_{m}$, it is computed as:
% \begin{equation*}
%     \left(\mathcal{X}\times_{m} U_{m}\right)(i_1,\ldots,i_{k-1},j,i_{k+1},\ldots,i_{K}) = \sum_{i_{k}=1}^{I_{k}} \mathcal{X}(i_1,\ldots,i_{k-1},i_{k},i_{k+1},\ldots,i_{K})\cdot U_{m}(j,i_{k})
% \end{equation*}
In effect, $\mathbf{A}$ and $\mathbf{B}$ reduce the dimension of each channel of $\mathcal{X}_{i}$ from $H\times W$ to $h\times w$ and one can rewrite \eqref{eq:tensor-contraction} equivalently as:
\begin{equation*}
    \mathbf{Z}_{i}^{(c)} = \mathbf{A}\mathbf{X}_{i}^{(c)}\mathbf{B}^{\top}, \quad c=1,2,\dots,C.
\end{equation*}
After \eqref{eq:tensor-contraction}, we then apply \eqref{eq:Baseline-GP} on $\mathcal{Z}_{i}$, as discussed later.

This formulation of tensor contraction can be found in a more general setting in tensor regression networks \cite{kossaifi2020tensor}, where tensor contraction can be applied to compress any tensors in a neural network. In our method, we envelope the tensor contraction operation within a tensor GP framework. Also, note that in \eqref{eq:tensor-contraction}, all channels share the same tensor contracting factors $\mathbf{A}$ and $\mathbf{B}$, which preserves the spatial consistency of different channels of the reduced-sized tensor $\mathcal{Z}$ for easier interpretation. Alternatively, one can replace the $\mathbf{I}_{C}$ in \eqref{eq:tensor-contraction} with an arbitrary $C\times C$ matrix $\mathbf{C}$, ending up with the full tensor contraction in \citet{kossaifi2020tensor}. We stick to \eqref{eq:tensor-contraction} for simplicity in this paper.

% This is a reasonable assumption to make since only when different channels share the same dimension reduction parameters will the latent tensors $\mathcal{Z}_{i}$ have element-wisely consistent interpretation.

\begin{figure*}[t]
    \centering
    \begin{subfigure}[b]{0.54\textwidth}
        \centering
        \includegraphics[width=0.98\textwidth]{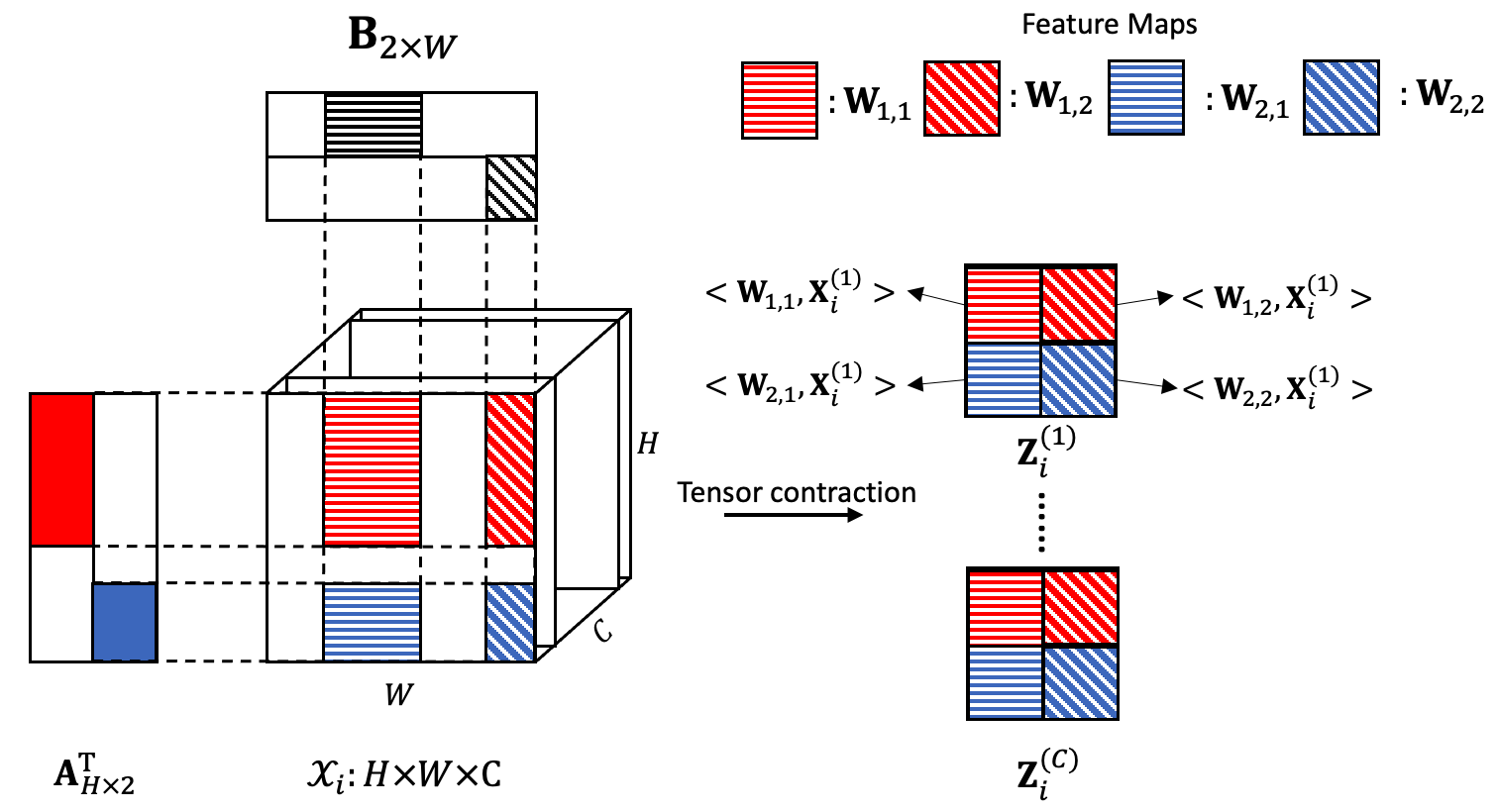}
        \caption{Tensor Contraction}
        \label{fig:tensor-contract}
    \end{subfigure}
    \begin{subfigure}[b]{0.44\textwidth}
        \centering
        \includegraphics[width=0.98\textwidth]{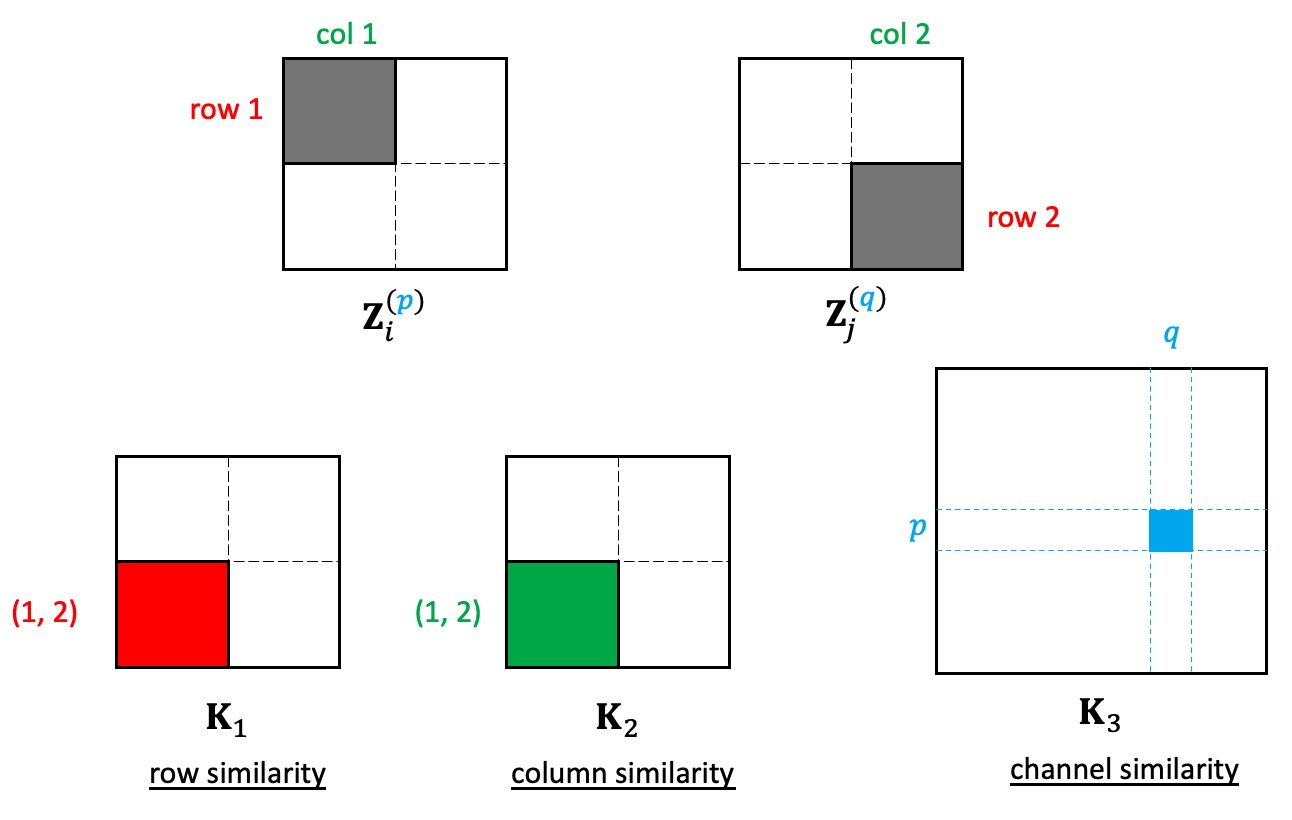}
        \caption{Multi-Linear Kernel Interpretation}
        \label{fig:tensor-GP}
    \end{subfigure}
    \caption{(a) Example of the tensor contraction step for tensor data $\mathcal{X}_{i}\in \mathbb{R}^{H\times W\times C}$ to its latent tensor $\mathcal{Z}_{i}\in \mathbb{R}^{2\times 2\times C}$. The tensor contracting factors $\mathbf{A}$, $\mathbf{B}$ are sparse (colored/dashed bands indicate nonzero elements) and they jointly extract features from $\mathbf{X}_{i}^{(1)}, \dots, \mathbf{X}_{i}^{(C)}$ with rank-1 feature maps $\left\{\mathbf{W}_{1,1},\mathbf{W}_{1,2},\mathbf{W}_{2,1},\mathbf{W}_{2,2}\right\}$. Each channel of $\mathbf{Z}_{i}^{(c)}$ has $2\times 2$ features, based on the inner product of every feature map with the channel data $\mathbf{X}_{i}^{(c)}$. (b) Example of the multi-linear kernel with a pair of latent tensor data $\left(\mathcal{Z}_{i}, \mathcal{Z}_{j}\right)$. Any pair of pixels in $\mathcal{Z}_{i}$ and $\mathcal{Z}_{j}$, e.g. $\mathbf{Z}_{i}^{(p)}(1,1)$ and $\mathbf{Z}_{j}^{(q)}(2,2)$ in the plot (colored in gray), are weighted by the product of their row similarity $\mathbf{K}_{1}(1,1)$ (red), column similarity $\mathbf{K}_{2}(2,2)$ (green) and channel similarity $\mathbf{K}_{3}(p,q)$ (blue), in the kernel function \eqref{eq:MLKernel} for defining the similarity of $\mathcal{Z}_{i}, \mathcal{Z}_{j}$. See \eqref{eq:single-channel-expvar} for a formulaic explanation.}
    \label{fig:tensor-GPST}
\end{figure*}

One can interpret the contracted tensor $\mathcal{Z}_{i}$ as the latent low-dimensional representation of the original tensor $\mathcal{X}_{i}$. Each $(s,t)^{\rm{th}}$ element of $\mathbf{Z}_{i}^{(c)}$ is constructed via a matrix inner product with a rank-1 ``feature map": $\mathbf{Z}_{i}^{(c)}(s,t) = \left<\boldsymbol{\alpha}_{s}^{\top}\boldsymbol{\beta}_{t}, \mathbf{X}_{i}^{(c)}\right>$, where $\boldsymbol{\alpha}_{s}$ and $\boldsymbol{\beta}_{t}$, the basis of the feature map, are the $s^{\rm th}$ and $t^{\rm th}$ rows of $\mathbf{A}$ and $\mathbf{B}$, respectively. We denote the feature map $\left(\boldsymbol{\alpha}_{s}^{\top}\boldsymbol{\beta}_{t}\right)$ as $\mathbf{W}_{s,t} \in \mathbb{R}^{H\times W}$. A visual explanation of the tensor contraction operation is shown in Figure \ref{fig:tensor-contract}. Note how elements of $\mathbf{Z}_{i}^{(c)}$ on the same row or column share the same feature map basis in $\mathbf{A}$ or $\mathbf{B}$.
% Since we preserve the tensor structure in the latent space, any slice of $\mathcal{Z}_{i}$ along the first or second mode share the same feature map basis $\boldsymbol{\alpha}_{s}$ or $\boldsymbol{\beta}_{t}$.

% Since we preserve the tensor structure of $\mathcal{X}_{i}$ in the latent space, any two elements in $\mathcal{Z}_{i}$ in the same row (column) share the same row (column) feature map in $\mathbf{A}$ ($B$).

Given the transformed tensor $\mathcal{Z}_{i} = g(\mathcal{X}_{i})$, we assume a GP prior for $\mathbf{y} = (y_{1}, y_{2}, \dots, y_{N})^{\top}$ given $\mathcal{Z}_{1}, \mathcal{Z}_{2}, \dots, \mathcal{Z}_{N}$ with a multi-linear kernel \cite{yu2018tensor}:
\begin{equation}
    y_{i} = h(\mathcal{Z}_{i}) + \epsilon_{i}, \quad h(\cdot) \sim \mbox{GP}\left(0, k(\cdot,\cdot)\right),\label{eq:MLGP-Z}
\end{equation}
where $k(\cdot,\cdot)$ is the multi-linear tensor kernel function:
\begin{equation}\label{eq:MLKernel}
    k(\mathcal{Z}_{i}, \mathcal{Z}_{j}) = \vect{\mathcal{Z}_{i}}^{\top} \left(\mathbf{K}_{3} \otimes \mathbf{K}_{2} \otimes \mathbf{K}_{1}\right) \vect{\mathcal{Z}_{j}}.
\end{equation}
The multi-linear kernel defines a similarity metric between pairs of tensor data. We provide an illustration of the multi-linear kernel in Figure \ref{fig:tensor-GP}. In this model, $\mathbf{K}_{1}\in \mathbb{R}^{h\times h}, \mathbf{K}_{2}\in \mathbb{R}^{w\times w}, \mathbf{K}_{3}\in \mathbb{R}^{C\times C}$ capture the mode-specific covariance structure.

% It has been shown that the multi-linear tensor GP can be interpreted as the probabilistic version of the low-rank tensor regression \cite{yu2018tensor}. Thus the model specified by (\ref{eq:MLGP-Z}) and (\ref{eq:MLKernel}), together with the tensor contraction in (\ref{eq:tensor-contraction}), can be rewritten as a tensor regression model with an extra dimension reduction step:
% \begin{align}
%     y_{i} & = \left<\mathcal{W}, \mathcal{Z}_{i}\right> + \epsilon_{i}\nonumber \\
%     & = \left<\mathcal{W}, \mathcal{X}_{i} \times_{1} A \times_{2} B \times_{3}\mathbf{I}_{C}\right> + \epsilon_{i} \label{eq:tensor_reg_equiv}
% \end{align}
% where $\mathcal{W} \in \mathbb{R}^{h\times w\times C}$ is the regression coefficient tensor with the same shape as the latent tensor $\mathcal{Z}_{i}$. This tensor regression model can be regarded as a two-layer tensor regression network with a tensor contraction layer and a tensor regression layer, as discussed in \citet{kossaifi2020tensor}. 

% In this paper, we investigate the Gaussian process version of the model specified by (\ref{eq:tensor-contraction}), (\ref{eq:MLGP-Z}) and (\ref{eq:MLKernel}), but we want to highlight that the GP version of the model is an alternative formulation of the two-layer tensor regression network with uncertainty quantification capability.

Combining (\ref{eq:tensor-contraction}), (\ref{eq:MLGP-Z}) and (\ref{eq:MLKernel}) together, 
% and using the property of vectorizing tensor mode product, 
our method essentially specifies the following tensor GP with a new kernel $\mathcal{K}(\cdot,\cdot)$:
\begin{align}
    y_{i} & = f(\mathcal{X}_{i}) + \epsilon_{i}, \quad f(\cdot) \sim \mbox{GP}\left(0, \mathcal{K}(\cdot,\cdot)\right) \label{eq:MLGP-X} \\
    \mathcal{K}\left(\mathcal{X}_{i}, \mathcal{X}_{j}\right) & = \vect{\mathcal{X}_{i}}^{\top}\left(\mathbf{K}_{3}\otimes \mathbf{K}_{2}^{*} \otimes \mathbf{K}_{1}^{*}\right)\vect{\mathcal{X}_{j}} \label{eq:MLGP-X-Kernel} \\
    & \mathbf{K}_{2}^{*} = \mathbf{B}^{\top}\mathbf{K}_{2}\mathbf{B}, \quad \mathbf{K}_{1}^{*} = \mathbf{A}^{\top}\mathbf{K}_{1}\mathbf{A}, \label{eq:MLGP-X-Kernel-Decomp}
\end{align}
and we call the framework \textbf{Tensor} \textbf{G}aussian \textbf{P}rocess with \textbf{S}patial \textbf{T}ransformation (\textbf{Tensor-GPST}), where $\mathbf{A}$ and $\mathbf{B}$ transform, in a bi-linear way, the spatial information contained in the imaging data.

Another way of expressing the model is via tensor regression \eqref{eq:tensor-regression} on the original tensor $\mathcal{X}$. Equivalently, we assume a Gaussian prior over $\mathcal{W}$:
\begin{align}
    \vect{\mathcal{W}} & \sim (\mathbf{I}_{C}\otimes \mathbf{B}\otimes \mathbf{A})^{\top}\vect{\mathcal{T}}, \mathcal{T} \in \mathbb{R}^{h\times w\times C}, \nonumber \\
    \vect{\mathcal{T}} & \sim \mathcal{N}\left(\mathbf{0}, \mathbf{K}_{3}\otimes \mathbf{K}_{2}\otimes \mathbf{K}_{1}\right), \label{eq:MLGP-prior}
\end{align}
which is similar to a tensor factor model \cite{chen2020semiparametric} coupled with a Gaussian factor with Kronecker-product covariance structure.

\subsection{Estimating Algorithm}\label{subsec:algo}
To estimate the model parameters of Tensor-GPST in (\ref{eq:MLGP-X})-(\ref{eq:MLGP-X-Kernel-Decomp}), including the tensor contracting factors $(\mathbf{A},\mathbf{B})$, the multi-linear kernel factors $(\mathbf{K}_{1}, \mathbf{K}_{2}, \mathbf{K}_{3})$, and the idiosyncratic noise variance $\sigma^2$, we minimize the negative marginal Gaussian log-likelihood $\ell(\mathbf{y}|\mathbf{A},\mathbf{B},\mathbf{K}_{1},\mathbf{K}_{2},\mathbf{K}_{3},\sigma)$:
\begin{equation}\label{eq:Negative-LL}
    \ell = \frac12 \log\left|\mathbf{K} + \mathbf{D}_{\sigma}\right| + \frac12 \mathbf{y}^{\top} \left(\mathbf{K} + \mathbf{D}_{\sigma}\right)^{-1}\mathbf{y} + \mbox{const.},
\end{equation}
where $\mathbf{K}$ is an $N\times N$ empirical kernel gram matrix computed using the kernel function (\ref{eq:MLGP-X-Kernel}) for all pairs of tensor data and $\mathbf{D}_{\sigma} = \sigma^{2}\mathbf{I}_{N}$.

To speed up the computation, we approximate each multi-linear kernel factor with a factorized form:
\begin{equation}
    \mathbf{K}_{1} = \mathbf{U}_{1}^{\top}\mathbf{U}_{1}, \mathbf{K}_{2} = \mathbf{U}_{2}^{\top}\mathbf{U}_{2}, \mathbf{K}_{3} = \mathbf{U}^{\top}_{3}\mathbf{U}_{3},
\end{equation}
where $\mathbf{U}_{1}\in \mathbb{R}^{r_{1}\times h}, \mathbf{U}_{2} \in \mathbb{R}^{r_{2}\times w}, \mathbf{U}_{3} \in \mathbb{R}^{r_{3}\times C}$. $\mathbf{U}_{1}, \mathbf{U}_{2}, \mathbf{U}_{3}$ are orthogonal matrices with $r_{1} \le h, r_{2} \le w, r_{3}\le C$. The tuning parameter is set as such that $r_{1}=h, r_{2}=w, r_{3}=C$ throughout the paper but can be set to smaller values to enforce a low-rank constraint. With the factorization assumption, one can decompose the gram matrix $\mathbf{K}$ as $\widetilde{\mathbf{U}}\widetilde{\mathbf{U}}^\top$, where:
\begin{equation*}
    \widetilde{\mathbf{U}} = \widetilde{\mathcal{X}}^{\top}\left(\mathbf{I}_{C}\otimes \mathbf{B}\otimes \mathbf{A}\right)^{\top} \left(\mathbf{U}_{3}\otimes \mathbf{U}_{2} \otimes \mathbf{U}_{1}\right)^{\top},
\end{equation*}
where $\widetilde{\mathcal{X}} = \left[\vect{\mathcal{X}_{1}}; \vect{\mathcal{X}_{2}}; \dots; \vect{\mathcal{X}_{N}}\right]$. The factorized form of $\mathbf{K}$ can simplify the computation of the gradients since one can invert the covariance matrix $(\mathbf{K} + \mathbf{D}_{\sigma})$ with the Woodbury identity, as shown in Appendix \ref{app:algorithm}. The computational complexity of the algorithm is thus reduced from the canonical $\mathcal{O}(N^{3})$ to $\mathcal{O}(N^{2}D)$, where $D=HWC$ is the dimension of the data tensor.

Since the tensor contracting factors $(\mathbf{A}, \mathbf{B})$ are extracting spatial features from each channel of $\mathcal{X}_i$, we assume that each spatial feature can be constructed from several spatially-contiguous regions for better interpretability. This leads us to the assumption that each feature map $\mathbf{W}_{s,t} = \boldsymbol{\alpha}_{s}^{\top}\boldsymbol{\beta}_{t}$ has certain degrees of spatial smoothness. We introduce the spatial smoothness assumption into our model via regularizing its anisotropic total variation norm $\|\mathbf{W}_{s,t}\|_{\text{TV}}$:
\begin{align*}
    \|\mathbf{W}_{s,t}\|_{\text{TV}} & = \sum_{i=1}^{H-1}\sum_{j=1}^{W}  \left|\mathbf{W}_{s,t}(i+1,j)-\mathbf{W}_{s,t}(i,j)\right| \\
    & + \sum_{i=1}^{H}\sum_{j=1}^{W-1}  \left|\mathbf{W}_{s,t}(i,j+1)-\mathbf{W}_{s,t}(i,j)\right|.
\end{align*}
A more general class of total variation norm penalty on tensor regression model coefficients can be found in \citet{wang2017generalized}. In Lemma \ref{thm:tv-norm-decomp}, we derive a simplified form of $\|\mathbf{W}_{s,t}\|_{\text{TV}}$, making the estimation of $\mathbf{A}$ and $\mathbf{B}$ easier.

\begin{lemma}\label{thm:tv-norm-decomp}
The anisotropic total variation (TV) norm on feature map $\{\mathbf{W}_{s,t}\}_{s=1,t=1}^{h,w}$ induces a fused-lasso \cite{tibshirani2005sparsity} penalization on $\mathbf{A}$ (and $\mathbf{B}$), namely:
\begin{equation}
    \sum_{s=1}^{h}\sum_{t=1}^{w} \|\mathbf{W}_{s,t}\|_{\mathrm{TV}} = \left\|\nabla_{x} \mathbf{B}\right\|_{1}\|\mathbf{A}\|_{1} + \|\mathbf{B}\|_{1}\left\|\nabla_{x} \mathbf{A}\right\|_{1} \label{eq:tv-fused-lasso},
\end{equation}
where $\nabla_{x}$ computes the horizontal gradient of a matrix, i.e. $\nabla_{x}\mathbf{A}_{m\times n}(i,j) = \mathrm{I}_{\{j\neq n\}}\left[\mathbf{A}(i,j+1) - \mathbf{A}(i,j)\right]$, and $\|\cdot\|_{1}$ is the elementwise 1-norm of a matrix. 
% We denote the above penalty as $\mathrm{R}(A,B)$ throughout the remainder of the paper.
\end{lemma}

% The proof is straightforward by plugging in the formula of each feature map $\mathbf{W}_{s,t}$. 
We leave the proof to Appendix \ref{app:lemma-1}.

The fused-lasso penalty penalizes the sparsity and smoothness of $\mathbf{A}$, weighted by the smoothness and sparsity of $\mathbf{B}$ and vice versa. Jointly, our estimating problem is attempting to minimize the following penalized negative log-likelihood:
\begin{equation}\label{eq:loss}
    L(\mathbf{y}|\mathbf{A},\mathbf{B},\mathbf{U}_{1:3},\sigma) = \ell(\mathbf{y}|\mathbf{A},\mathbf{B},\mathbf{U}_{1:3},\sigma) + \lambda \mathrm{R}(\mathbf{A},\mathbf{B}),
\end{equation}
where $\mathrm{R}(\mathbf{A},\mathbf{B}) = \left\|\nabla_{x} \mathbf{B}\right\|_{1}\|\mathbf{A}\|_{1} + \|\mathbf{B}\|_{1}\left\|\nabla_{x} \mathbf{A}\right\|_{1}$ and $\mathbf{U}_{1:3}$ is the collection of $\mathbf{U}_{1},\mathbf{U}_{2},\mathbf{U}_{3}$.

The total variation penalty can create feature maps with sharp edges and leads to sparsity for interpretation. In the estimating algorithm, we use proximal gradient descent to estimate the tensor contracting factors $(\mathbf{A}, \mathbf{B})$ and cyclically update the parameters in the order of: $\mathbf{A} \rightarrow \mathbf{B} \rightarrow (\mathbf{U}_{1},\mathbf{U}_{2},\mathbf{U}_{3}) \rightarrow \sigma \rightarrow \mathbf{A} \rightarrow \dots$. The fused-lasso penalty over $\mathbf{A}$ and $\mathbf{B}$ makes the proximal step a well-defined \textit{fused lasso 1-D signal approximation} problem \cite{friedman2007pathwise}. Specifically, at the $(i+1)^{\rm{th}}$ iteration, we first propose a gradient descent update for $\mathbf{A}$, denoted as $\widehat{\mathbf{A}}^{(i+\frac{1}{2})}$, with stepsize $\eta_{i}$. The final updated value for $\mathbf{A}$, i.e. $\widehat{\mathbf{A}}^{(i+1)}$, is the minimizer of the proximal step:
\begin{align*}
    \widehat{\mathbf{A}}^{(i+1)} & = \mathbf{prox}_{\text{TV}}\left(\widehat{\mathbf{A}}^{(i+\frac{1}{2})}\right) \\
    & = \argmin_{\mathbf{A}} \left\{\frac{1}{2\eta_{i}}\left\|\mathbf{A} - \widehat{\mathbf{A}}^{(i+\frac{1}{2})}\right\|^{2}_{\mathrm{F}} + \lambda \mathrm{R}(\mathbf{A},\widehat{\mathbf{B}}^{(i)})\right\},
\end{align*}
which can be easily solved by first solving the minimization without the $\ell_{1}$-penalty on $\mathbf{A}$ and then apply a soft-thresholding operator to obtain the exact minimizer (see Proposition 1 of \citet{friedman2007pathwise} for the justification). The same procedure applies when one updates $\mathbf{B}$. We summarize the outline of the estimating algorithm in Algorithm \ref{algorithm} and provide the details of the derivation of gradients and the proximal step in Appendix \ref{app:algorithm}.

Since any pair of $(\mathbf{A},\mathbf{B})$ can be re-scaled by a constant $c_1$ such that: $(\mathbf{B}\otimes \mathbf{A}) = (c_{1}^{-1}\mathbf{B})\otimes(c_{1}\mathbf{A})$, we re-scale the norm of $(\widehat{\mathbf{A}}^{(i)},\widehat{\mathbf{B}}^{(i)})$ after each iteration to ensure that there is no scaling identifiability issue for the tensor contraction operation.

We do not enforce the orthonormality of $\mathbf{U}_{1}, \mathbf{U}_{2}, \mathbf{U}_{3}$, but a good initialization can still obtain reasonable approximations according to \citet{yu2018tensor}. To give a warm start of the model parameters, one can consider solving a tensor regression problem and a tucker decomposition problem subsequently, as inspired by \eqref{eq:MLGP-prior}:
\begin{equation}\label{eq:tensor-reg-warm-start}
    \min_{\substack{\mathcal{T}\in\mathbb{R}^{h\times w\times C} \\ \mathbf{A}\in\mathbb{R}^{h\times H} \\ \mathbf{B}\in\mathbb{R}^{w\times W}}} \sum_{i=1}^{N} \left(y_{i} - \left<\mathcal{X}_{i}, \mathcal{T}\times_{1} \mathbf{A}^{\top}\times_{2} \mathbf{B}^{\top}\right>\right)^{2},
\end{equation}
\begin{equation}\label{eq:tucker-warm-start}
    \min_{\substack{\mathcal{S}\in\mathbb{R}^{r_{1}\times r_{2}\times r_{3}} \\ \mathbf{U}_{1}, \mathbf{U}_{2}, \mathbf{U}_{3}}} \left\|\widehat{\mathcal{T}} - \mathcal{S}\times_{1}\mathbf{U}_{1}^{\top}\times_{2} \mathbf{U}_{2}^{\top}\times_{3} \mathbf{U}_{3}^{\top}\right\|^{2}.
\end{equation}
One obtains $\widehat{\mathbf{A}}^{(0)},\widehat{\mathbf{B}}^{(0)}$ from \eqref{eq:tensor-reg-warm-start} and $\widehat{\mathbf{U}}_{1:3}^{(0)}$ from \eqref{eq:tucker-warm-start}.

\begin{algorithm}
\caption{Alternating Proximal Gradient Descent Algorithm for Tensor-GPST Estimation}\label{algorithm}
\begin{algorithmic}
\STATE Initialize $\widehat{\mathbf{A}}^{(0)}, \widehat{\mathbf{B}}^{(0)}, \widehat{\mathbf{U}}_{1}^{(0)}, \widehat{\mathbf{U}}_{2}^{(0)}, \widehat{\mathbf{U}}_{3}^{(0)}, \widehat{\sigma}^{(0)}$ randomly.
\STATE Set iteration counter $i\leftarrow 0$.
\WHILE {not converge and $i \le $max-iter}
\STATE $\widehat{\mathbf{A}}^{(i+\frac12)} \leftarrow \widehat{\mathbf{A}}^{(i)}-\eta_{i}\nabla_{\mathbf{A}}\ell(\mathbf{y}|\widehat{\mathbf{A}}^{(i)}, \widehat{\mathbf{B}}^{(i)}, \widehat{\mathbf{U}}_{1:3}^{(i)}, \widehat{\sigma}^{(i)})$.
\STATE $\widehat{\mathbf{A}}^{(i+1)} \leftarrow \mathbf{prox}_{\text{TV}}(\widehat{\mathbf{A}}^{(i+\frac12)})$. \hfill // Fused-Lasso
\STATE $\widehat{\mathbf{B}}^{(i+\frac12)} \leftarrow \widehat{\mathbf{B}}^{(i)}-\eta_{i}\nabla_{\mathbf{B}}\ell(\mathbf{y}|\widehat{\mathbf{A}}^{(i+1)}, \widehat{\mathbf{B}}^{(i)}, \widehat{\mathbf{U}}_{1:3}^{(i)}, \widehat{\sigma}^{(i)})$.
\STATE $\widehat{\mathbf{B}}^{(i+1)} \leftarrow \mathbf{prox}_{\text{TV}}(\widehat{\mathbf{B}}^{(i+\frac12)})$. \hfill // Fused-Lasso
\STATE Re-scale $\widehat{\mathbf{A}}^{(i+1)}, \widehat{\mathbf{B}}^{(i+1)}$ s.t. $\|\widehat{\mathbf{A}}^{(i+1)}\|_{\mathrm{F}}=1$.
\FOR {j=1:3}
\STATE $\mathbf{G}_{j} \leftarrow \nabla_{\mathbf{U}_{j}}\ell(\mathbf{y}|\widehat{\mathbf{A}}^{(i+1)}, \widehat{\mathbf{B}}^{(i+1)}, \widehat{\mathbf{U}}_{-j}^{(i)}, \widehat{\sigma}^{(i)})$.\footnotemark
\STATE $\widehat{\mathbf{U}}_{j}^{(i+1)} \leftarrow \widehat{\mathbf{U}}_{j}^{(i)}-\eta_{i}\mathbf{G}_{j}$.
\ENDFOR
% \STATE Re-scale $\left(\widehat{\mathbf{U}}_{1}^{(i)},\widehat{\mathbf{U}}_{2}^{(i)},\widehat{\mathbf{U}}_{3}^{(i)}\right)$ s.t. $\|\widehat{\mathbf{U}}_{1}^{(i)\top}\widehat{\mathbf{U}}_{1}^{(i)}\|_{F}=\|\widehat{\mathbf{U}}_{2}^{(i)\top}\widehat{\mathbf{U}}_{2}^{(i)}\|_{F}=1$. 
\STATE $t \leftarrow \nabla_{\sigma}\ell(\mathbf{y}|\widehat{\mathbf{A}}^{(i+1)}, \widehat{\mathbf{B}}^{(i+1)}, \widehat{\mathbf{U}}_{1:3}^{(i+1)}, \widehat{\sigma}^{(i)})$.
\STATE $\widehat{\sigma}^{(i+1)} \leftarrow \widehat{\sigma}^{(i)} - \eta_{i}t$. 
\STATE $i \leftarrow i + 1$
\ENDWHILE
\STATE \textbf{Output}: $\widehat{\mathbf{A}}^{(i)}, \widehat{\mathbf{B}}^{(i)}, \widehat{\mathbf{U}}_{1:3}^{(i)}, \widehat{\sigma}^{(i)}$
\end{algorithmic}
\end{algorithm}
\footnotetext{We use $\mathbf{U}_{-j}$ to denote the collection of $\mathbf{U}_{1},\mathbf{U}_{2},\mathbf{U}_{3}$ but exclude $\mathbf{U}_{j}$.}

%In the algorithm, we re-scale the norm of $(\mathbf{A},\mathbf{B})$ to ensure parameter scale identifiability within the tensor contraction step. 

\subsection{Convergence Analysis}\label{subsec:theory}
In this subsection, we provide the convergence analysis of Algorithm \ref{algorithm}. Theorem \ref{thm:conv-rate} provides the upper bound of the loss function \eqref{eq:loss}, evaluated at the estimators output by the algorithm, with respect to its global minimum. We show that the total variation penalty and the alternating proximal gradient descent introduce extra gaps between the achieved loss and its global minimum.

\begin{theorem}\label{thm:conv-rate}
Given the loss function $L(\cdot)$ in \eqref{eq:loss}, assume that the negative log-likelihood $\ell(\cdot)$ is convex for any of the four parameter blocks: $\{\mathbf{A}\}, \{\mathbf{B}\}, \{\mathbf{U}_{1:3}\}, \{\sigma\}$, with the other three blocks being fixed, and the gradients of $\ell(\cdot)$ are Lipschitz continuous with Lipschitz constant: $M_{\mathbf{A}}, M_{\mathbf{B}}, M_{\mathbf{U}}, M_{\sigma}$, respectively. Then with a constant learning rate $\alpha \le 1/\max\{M_{\mathbf{A}}, M_{\mathbf{B}}, M_{\mathbf{U}}, M_{\sigma}\}$, the alternating proximal gradient descent algorithm in Algorithm \ref{algorithm} leads to the following upper bound on the loss function $L(\cdot)$:
\begin{align*}
    & 4\left(K+1\right)\left[L(\widehat{\boldsymbol{\theta}}^{(K+1)}) - L(\boldsymbol{\theta}^{*})\right]  \le \frac{\delta^{(0)}}{2\alpha} \\
    & + \sum_{k=0}^{K} h_{\lambda}(\widehat{\mathbf{A}}^{(k+1)}-\mathbf{A}^{*}, \widehat{\mathbf{B}}^{(k+1)}-\mathbf{B}^{*}, \widehat{\mathbf{B}}^{(k)}-\mathbf{B}^{*}) \\
    & + \frac{1}{2\alpha}\sum_{k=0}^{K}\tau\left(\widehat{\boldsymbol{\theta}}^{(k+1)}, \widehat{\boldsymbol{\theta}}^{(k)}, \boldsymbol{\theta}^{*}\right), \numberthis \label{eq:Error-Upper-Bound}
\end{align*}
where $\widehat{\boldsymbol{\theta}}^{(k)} = \left\{\widehat{\mathbf{A}}^{(k)},\widehat{\mathbf{B}}^{(k)},\widehat{\mathbf{U}}^{(k)}_{1:3}, \widehat{\sigma}^{(k)}\right\}$ and $\boldsymbol{\theta}^{*}$ is the global minimizer of $L(\cdot)$. $\delta^{(0)}$ is the squared $\ell_{2}$ initialization error, $h_{\lambda}(\cdot) \ge 0$ is the total-variation gap (TV-gap) and $\tau(\cdot) \ge 0$ is the alternating descent gap (ALT-gap). $K$ is the total number of iterations. As a result, if one has any three blocks of parameters fixed at their global minima, the remaining block will converge to its global minima at the rate of $\mathcal{O}(1/K)$, which echoes the convergence rate of (proximal) gradient descent.
\end{theorem}

We leave the proof to Appendix \ref{app:proof-conv-rate} and make a few remarks.

\begin{remark}
As $\widehat{\mathbf{A}}^{(k)}$ \textrightarrow $\mathbf{A}^{*}$, $\widehat{\mathbf{B}}^{(k)}$ \textrightarrow $\mathbf{B}^{*}$, one has $h_{\lambda}(\cdot)$ \textrightarrow $0$. The TV-gap is incurred because we alternatively update $\mathbf{A}$ and $\mathbf{B}$, and using the current iteration's estimate of $\mathbf{A}$ (or $\mathbf{B}$) for updating $\mathbf{B}$ (or $\mathbf{A}$) with the total variation penalty leads to extra errors. See the definition of $h_{\lambda}(\cdot)$ in \eqref{eq:TV-gap-form}.
\end{remark}

\begin{remark}
As $\widehat{\boldsymbol{\theta}}^{(k)}$ \textrightarrow $\boldsymbol{\theta}^{*}$, $\tau(\cdot)\rightarrow 0$. The ALT-gap $\tau(\cdot)$ arises because we use the current iteration's estimate for all but one block of parameters to estimate the gradient of the block of interest. If the algorithm terminates at a local minima, the non-vanishing TV-gap and ALT-gap leads to a non-zero gap for the achieved loss from the global minimum. See the definition of $\tau(\cdot)$ in \eqref{eq:ALT-gap-form}.
\end{remark}

\begin{remark}
Tensor regression models with Tucker-type low-rankness have non-convex negative-likelihood function $\ell(\cdot)$ \cite{li2018tucker}. But conditioning on all but one block of parameter, $\ell(\cdot)$ is convex for each individual block. We do not verify the convexity of $\ell(\cdot)$ in our particular model due to the complexity of the kernel function. Empirically, as we show in Figure \ref{fig:loss-function-example} in Appendix \ref{app:proof-conv-rate} and also in \citet{yu2018tensor}, such alternating gradient descent algorithm works well with the optimization problem and the loss function decays at the rate of $\mathcal{O}(1/K)$.
\end{remark}

% \begin{remark}
% In Theorem \ref{thm:conv-rate}, if any three blocks of parameters are fixed at the global minimum, and one only updates the remaining block, the loss function converges to the global minimum $L(\boldsymbol{\theta}^{*})$ at the rate of $\mathcal{O}\left(1/K\right)$, which is a classical result of (proximal) gradient descent.
% \end{remark}

\section{Experiments}
In this section, we validate our method via both simulation studies and an application to an astrophysics dataset for solar flare forecasting. We also compare our method against other benchmark tensor regression models. In particular, we are interested in applications to imaging data where the predictive signals appear in different channels and various locations within a channel. Such patterns are common in astrophysical imaging data where the solar flare precursors could appear in the images collected by the astrophysical telescopes at various frequencies and the locations of the precursors could be random within an image channel.
% Such patterns are common in scientific datasets where it is hard to define a ``canonical" position of different imaging, and different samples with similar scalar labels can have similar patterns up to a shift in location and a change in channel.

\subsection{Simulation Study}
\begin{figure}[htb]
    \centering
    \includegraphics[height=9.5cm]{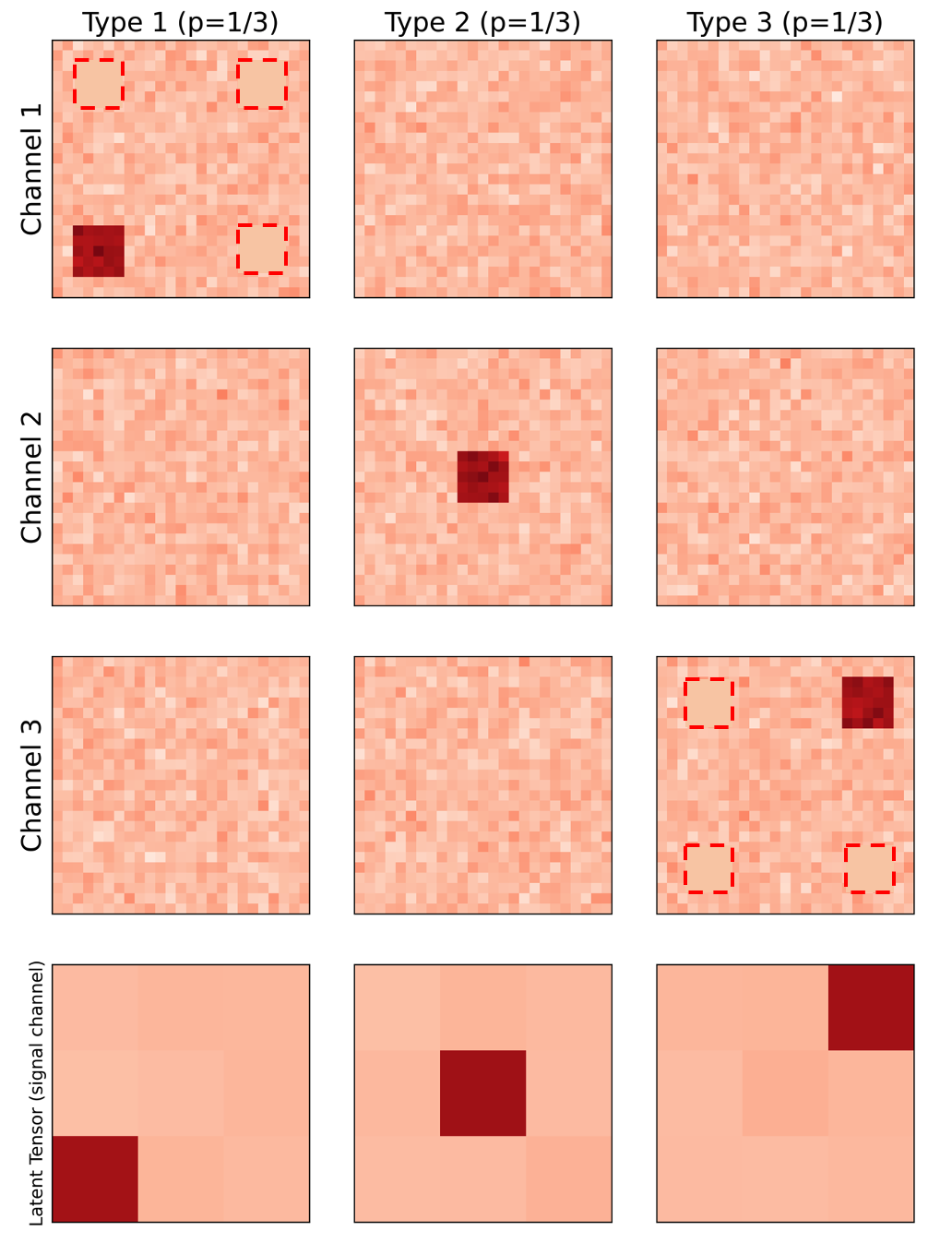}
    \caption{Three types of the simulated tensor data ($\mathcal{X}_{i} \in \mathbb{R}^{25\times 25\times 3}$). Each column is a type (Type 1,2,3) and every sample has equal probability of being one of the three types. Each row (row $1$-$3$) is a data channel (channel 1,2,3). Type 1, 2 and 3 have their \textit{signal} channel in channel 1, 2 and 3, respectively. But the location of the $5\times 5$ signal block is positioned differently. Type 2 has the signal fixed at the center, while type 1 and 3 has the signal placed, with equal probability, in one of the four corners (dashed block shows the other three possible locations). Samples shown are one realization of the simulation. The latent tensor $\mathcal{Z}$'s signal channel is shown at the bottom. See details in Appendix \ref{app:simulation}.}
    \label{fig:simdata}
\end{figure}

We simulate a tensor dataset $\left\{\mathcal{X}_{i}\right\}_{i=1}^{N}$ with each $\mathcal{X}_{i}$ having $3$ channels of size $25\times 25$. For each $25\times 25\times 3$ tensor data $\mathcal{X}_{i}$, we randomly pick one of the three channels as the \textit{signal} channel, with equal probability, and the remaining two channels as the \textit{noise} channels. The noise channel contains i.i.d. pixels from $\mathcal{N}(0,0.3)$, and the signal channel uses the same background noise distribution except having a $5\times 5$ \textit{signal} block that contains i.i.d. pixels from $\mathcal{N}(4,0.3)$. The location of the $5\times 5$ block is fixed at the center of the $25\times 25$ image if channel $2$ is the signal channel (see Type 2 in Figure \ref{fig:simdata}), and is randomly picked at one of the four corners (top-left, top-right, bottom-left, bottom-right) if channel 1 or 3 is the signal channel (see Type 1 and 3 in Figure \ref{fig:simdata}).

We simulate the tensor contracting factors $\mathbf{A},\mathbf{B} \in \mathbb{R}^{3\times 25}$ with a banded structure, leading to a $3\times 3\times 3$ contracted tensor $\mathcal{Z}$, such that $\mathbf{A}$ and $\mathbf{B}$ are extracting features from the $5\times 5$ blocks with \textit{signal}, see the bottom of Figure \ref{fig:simdata} for the example of the contracted tensor $\mathcal{Z}$. The multi-linear kernel setup and the generating process of the regression labels $\{y_{i}\}_{i=1}^{N}$ are detailed in Appendix \ref{app:simulation}. Generally, channel 2 is simulated such that it is negatively correlated with channels 1 \& 3, and channels 1 \& 3 are nearly perfectly correlated. As a result, Type 1 \& 3 tensors have similar regression labels and differ from those of Type 2.

With the simulation setups, we compare our model against these baseline tensor regression models: Tensor-GP (\textbf{GP}) \cite{yu2018tensor}, $\ell_{2}$-regularized tensor regression with CANDECOMP/PARAFAC (CP) tensor rank constraints \cite{guo2011tensor} (\textbf{CP}), $\ell_{2}$-regularized tensor regression with coefficient tensor following a Tucker decomposition \cite{li2018tucker} (\textbf{Tucker}). In order to check the sensitivity of the choice of the kernel, we also fit the GP model to the vectorized tensor data with a squared-exponential kernel (\textbf{SE}). To showcase the effectiveness of tensor contraction, we also consider fitting a model with a tensor contraction step followed by a GP with squared-exponential kernel for the vectorized, reduced-sized tensor (\textbf{SE+TC}). For simplicity, we implement \textbf{SE+TC} by training the model, without the total variation penalty, in an end-to-end fashion with the \texttt{GPyTorch} and \texttt{Tensorly-Torch} packages in \texttt{Python}. Both models involving the SE kernel have automatic relevance determination (ARD) length scales \cite{bishop2006pattern}.

% For our own Tensor-GPST (\textbf{GPST}) model, we train the model by picking the tuning parameter lambda with $5$-fold cross validation.
% We consider three variations of our model: model with hard constraints on the sparsity of $(A, B)$ based on the ground truth knowledge (\textbf{GPST-Hard}), model with low ($\lambda = 0.1$) total variation penalty (\textbf{GPST-low}), and model with high ($\lambda = 1.0$) total variation penalty (\textbf{GPST-high}). 
We simulate the data with size $N \in \{200, 500\}$ and use $75\%$ for training and $25\%$ for testing and compare the rooted-mean-squared-error (RMSE) on both training and testing across all models above as well as our own Tensor-GPST model (\textbf{GPST}). We set the latent tensor dimension as $3\times 3\times 3$ for \textbf{GPST} and \textbf{SE+TC} and the rank for $\mathbf{K}_{1},\mathbf{K}_{2},\mathbf{K}_{3}$ of \textbf{GP} as $3$ and the CP rank as $9$ for \textbf{CP} and the multi-linear rank as $3\times 3\times 3$ for \textbf{Tucker} such that the low-rankness is comparable across all methods. We select the regularization tuning parameter for all models with hyperparameters by $5$-fold cross validation. The simulation experiment is iterated 10 times and the testing RMSE is shown in Table \ref{tab:test_RMSE_simulation}.

\begin{table}[htb]
    \centering
    \caption{Test prediction RMSE for simulated Data for various tensor regression models. $95\%$ confidence interval after $\pm$. Results are based on 10 repeated runs.}
    \begin{tabular}{|c|c|c|}
    \hline
    Model  & $N=200$ & $N=500$ \\\hline
    \textbf{GP} & 0.728$\pm$0.125 & 0.664$\pm$0.131 \\ 
    \textbf{CP} & 0.550$\pm$0.100 & 0.548$\pm$0.054 \\
    \textbf{Tucker} & 0.589$\pm$0.206 & 0.568$\pm$0.107 \\
    \textbf{SE} & 2.504$\pm$2.672 & 3.275$\pm$4.204 \\
    \textbf{SE+TC} & 0.627$\pm$0.169 & 0.587$\pm$0.098 \\
    \textbf{GPST} (Our Method) & 0.578$\pm$0.107 & 0.552$\pm$0.076 \\
    \hline
    \end{tabular}
    \label{tab:test_RMSE_simulation}
\end{table}

The Tensor-GP (\textbf{GP}) method has relatively worse performance on the testing set compared to other low-rank tensor regression methods such as \textbf{CP} and \textbf{Tucker}. Our method, namely \textbf{GPST}, achieves similar performance to the low-rank tensor regression methods (not statistically significantly worse). The GP with vectorized tensor data and squared-exponential kernel, namely \textbf{SE}, performs extremely poorly, which reveals the fact that by vectorizing tensor data one loses the essential structural information of the data. This result necessitates the choice of kernel that is suitable for tensor data, such as the tensor GP. After adding an extra tensor contraction step, the GP with squared-exponential kernel (i.e. \textbf{SE+TC}) performs relatively close to the low-rank tensor regression methods as well as our \textbf{GPST} and is better than the tensor GP. This further suggests that regardless of the kernel choice, the tensor contraction step can boost the performance of GP regression models with tensor covariates. Effectively, the tensor contraction step extracts useful features from the original tensor data for regression, so even if one vectorizes the reduced-sized tensor, one does not lose as much information as the case where tensor contraction is not being used. Finally, we note that with a large sample size ($N=500$), the prediction RMSE of the test set gets smaller for all methods but \textbf{SE}.

To make further comparisons of the variants of different Gaussian Process models listed in Table \ref{tab:test_RMSE_simulation} on their ability to quantify the uncertainties of the predictions made, we compare these GP models' mean standardized log loss (MSLL) \cite{williams2006gaussian}, as defined below:
\begin{equation*}\label{eq:MSLL}
    \text{MSLL} = \frac{1}{N_{\text{test}}}\sum_{i=1}^{N_{\text{test}}} \left\{\frac12 \log\left(2\pi\widehat{\sigma}^2\right) + \frac{(y_{i} - \widehat{y}_{i})^2}{2\widehat{\sigma}^2}\right\},
\end{equation*}
where $\widehat{y}_{i}$ is the predicted label for the $i^{\rm{th}}$ testing sample and $\widehat{\sigma}$ is the estimated standard deviation of the noise term. Generally speaking, a smaller MSLL indicates a better testing prediction. We list the testing set MSLL for \textbf{GP}, \textbf{SE}, \textbf{SE+TC} and \textbf{GPST} in Table \ref{tab:test_MSLL_simulation}.

\begin{table}[htb]
    \centering
    \caption{Test Mean Standardized Log Loss (MSLL) for the 4 variants of GP models. $95\%$ confidence interval after $\pm$. Results are based on 10 repeated runs.}
    \begin{tabular}{|c|c|c|}
    \hline
    Model & $N=200$ & $N=500$  \\\hline
    \textbf{GP} & 1.162$\pm$0.439 & 1.092$\pm$0.421 \\ 
    \textbf{SE} & 2.457$\pm$1.898 & 2.999$\pm$3.167 \\
    \textbf{SE+TC} & 0.972$\pm$0.214 & 0.919$\pm$0.123 \\
    \textbf{GPST} & $\mathbf{0.882\pm0.201}$ & $\mathbf{0.835\pm0.156}$ \\
    \hline
    \end{tabular}    
    \label{tab:test_MSLL_simulation}
\end{table}

The result reveals that our method has statistically significantly smaller MSLL, as indicated by a one-sided paired t-test, compared to the other methods under both sample sizes. Also, the models with tensor contraction, including \textbf{SE+TC} and \textbf{GPST}, have smaller MSLL compared to their counterparts without tensor contraction, which further suggests that tensor contraction can be helpful for reducing the errors made by GP models with tensor data.

\begin{figure}[htb]
    \centering
    \includegraphics[height=10cm]{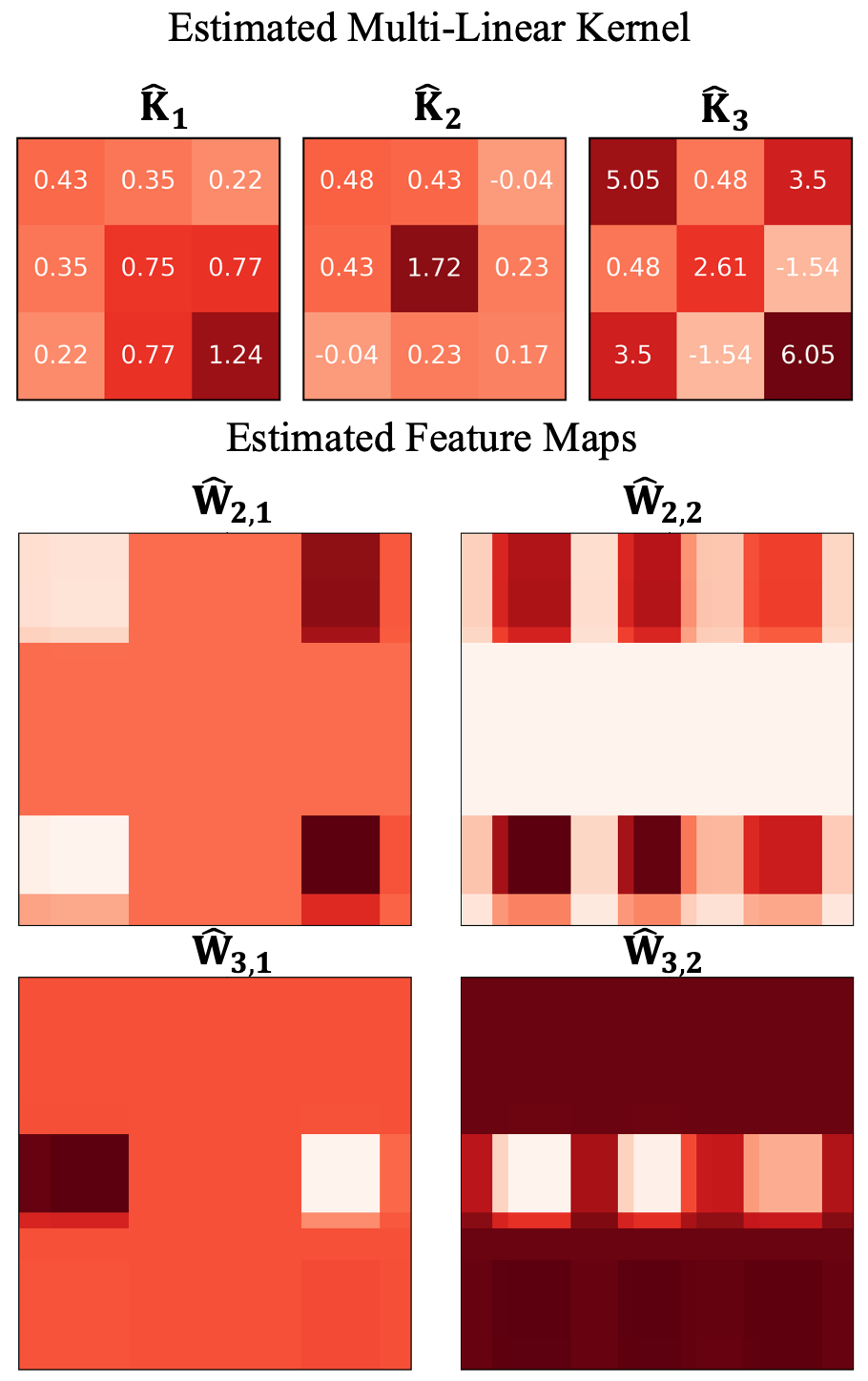}
    \caption{Estimated kernels (top) and non-zero feature maps (bottom) by \textbf{GPST} with $\lambda=1.0$ for one random simulation dataset.}
    \label{fig:simulation-estimates}
\end{figure}

The estimators of the multi-linear kernel factors $\mathbf{K}_{1},\mathbf{K}_{2},\mathbf{K}_{3}$ and the feature maps of the Tensor-GPST model with $\lambda=1.0$ for one random simulation dataset are visualized in Figure \ref{fig:simulation-estimates}. One can see that the feature map $\widehat{\mathbf{W}}_{2,2}$ and $\widehat{\mathbf{W}}_{3,2}$ capture the corner and center blocks, and the covariances between the two feature maps are also high, as suggested by $\widehat{\mathbf{K}_{1}}(2,3)=0.77$ and $\widehat{\mathbf{K}_{2}}(2,2)=1.72$. Channels 1 \& 3 have high covariances ($\widehat{\mathbf{K}_{3}}(1,3)=3.5$), indicating that they share similar ``corner signal" patterns and coincides with our ground truth setup (see Figure \ref{fig:Simulation-AB-Kernel} for the ground truth of $\mathbf{K}_{3}$).

Overall, the simulation experiments convey two messages:
\begin{itemize}
    \item Adding the tensor contraction step leads to better regression performances when the signals have low-rank structures, robust to the choice of kernel, and the performance is similar to other low-rank tensor regression methods such as \textbf{CP} and \textbf{Tucker}.
    \item The anisotropic total variation penalty, though may not fully recover the underlying sparsity of the tensor contracting factors, can improve the regression performance of Tensor-GPST and also provides more direct interpretations.
\end{itemize}

The inferior performance of Tensor-GP (\textbf{GP}), however, is not suggesting that it is an inferior version of GP when dealing with tensor data, as we have demonstrated by comparing it against the GP with squared-exponential kernel (\textbf{SE}). The simulation pattern in Figure \ref{fig:simdata} contains randomness of the signal, making it more beneficial to extract features first using feature maps that cover multiple areas. Directly modeling the covariance structures among all pixels can be difficult in such scenarios.

\subsection{Application to Solar Flare Forecasting}
A solar flare is an intense localized eruption of electromagnetic radiation in the Sun's atmosphere. Solar flares with high-energy radiation emission can strongly impact the Earth's space weather and potentially interfere the radio communication of the Earth. Recent works on solar flare forecasting \cite{bobra2015solar,chen2019identifying,wang2020predicting,jiao2020solar,sun2022predicting} have demonstrated the effectiveness of using machine learning algorithms for forecasting flares, using either multivariate time-series data in the form of physical parameters or imaging data provided by the Solar Dynamics Observatory (SDO)/Helioseismic and Magnetic Imager (HMI) \citep{Sherrer:2012} and SDO/Atmospheric Imaging Assembly (AIA) \citep{Lemen:2012}. It has been shown that these imaging data have low-dimensional representations that contain flare discriminating signals \cite{sun2021improved}. Our methodology makes the astrophysical interpretation more accessible as compared to the previous deep learning approaches in that our model has a much shallower structure with only a feature extraction layer and a regression layer.

Here, we consider the specific problem of forecasting the intensity of a solar flare. In our dataset, we have $1,329$ flare samples from year $2010$ to $2018$, consisting of a total of $479$ M-class and X-class flares and $850$ B-class flares. The class of a flare is determined by the X-ray peak brightness in the range of 1-8$\angstrom$. The B-class flare has its brightness within $10^{-7}\sim 10^{-6} \mbox{W/m}^{2}$, which is considered weak and barely harmful, while the minimum M-class and X-class flares have brightness above $10^{-5}$ and $10^{-4} \mbox{W/m}^{2}$, respectively. These more energetic flares are capable of heating and ionizing the the upper atmosphere, resulting in brief radio blackouts and increased satellite drag. We collect the AIA-HMI imaging data for each flare, $1$ hour prior to its peak, and each flare data is a 10-channel tensor data of size $50\times 50\times 10$, where spatial dimensions are binned down by roughly a factor of 10. We leave the data preprocessing steps and the astrophysical background to Appendix \ref{app:flare-dataset}.

Our goal here is to utilize the 10-channel tensor data $\mathcal{X}_{i}$ to predict the flare intensity $y_{i}$ and find the discriminating factors for M/X-class and B-class flares. We randomly split our dataset into a $75\%$ training set ($359 \mbox{ M/X}/637 \mbox{ B}$) and a $25\%$ testing set ($120 \mbox{ M/X}/213 \mbox{ B}$), and centering after log-transforming the flare intensity such that the B-class flare has $y_{i} \le -0.5$ and M/X-class flare has $y_{i} \ge 0.5$.

We report the solar flare intensity prediction result across four different models: Tensor-GP (\textbf{GP}), Tensor-GPST (\textbf{GPST}), tensor regression with CP rank constraints (\textbf{CP}) and tensor regression with Tucker decomposition form (\textbf{Tucker}). The hyperparameters are set such that the models have the same latent dimensionality ($3\times3\times 3$) or the rank ($9$ for \textbf{CP} and $3\times3\times 3$ for \textbf{Tucker}) of the regression coefficients. The metrics used are rooted mean-squared error (RMSE), R-squared and MSLL. Additionally, we consider transforming the regression model to a binary classification model by thresholding the prediction at $0.0$ such that any $\widehat{y}_{i} \ge 0$ indicates an M/X-class flare and any $\widehat{y}_{i} < 0$ indicates a B-class flare. Then we evaluate the resulting binary classification model with the True Skill Statistics (TSS)\footnote{True Skill Statistics is defined as: $\mbox{TSS} = \mbox{TP/(TP+FN)} - \mbox{FP/(FP+TN)}$, where TP, TN, FP, FN represents true positive, true negative, false positive and false negative in the confusion matrix.}. A skillful binary classifier for weak vs. strong solar flare is desirable for operational use. Results on the training and testing set are summarized in Table \ref{tab:flare-results}, with 10 random train/test splits. 

\begin{table*}[t]
    \centering
    \caption{Solar flare intensity regression performance on the training and testing sets for four tensor regression models. Results based on 10 random splits and $95\%$ confidence intervals are provided after $\pm$.}
    \resizebox{\textwidth}{!}{
    \begin{tabular}{|c|cccc|cccc|}
    \hline
    \multirow{2}{*}{Model} & \multicolumn{4}{|c|}{Training ($75\%$ of the samples)} & \multicolumn{4}{|c|}{Testing ($25\%$ of the samples)} \\\cline{2-9}
    & RMSE & $\mathrm{R}^{2}$ & MSLL & TSS &  RMSE & $\mathrm{R}^{2}$ & MSLL & TSS \\\hline
    \textbf{GP} & 0.646$\pm$0.019 & 0.336$\pm$0.044 & 1.028$\pm$0.134 & 0.466$\pm$0.039 & 0.772$\pm$0.239 & 0.182$\pm$0.114 & 1.138$\pm$0.085 & 0.362$\pm$0.159 \\
    \textbf{CP} & $\mathbf{0.564\pm0.035}$ & $\mathbf{0.501\pm0.077}$ & $-$ & $\mathbf{0.625\pm0.069}$ & 0.706$\pm$0.051 & 0.230$\pm$0.078 & $-$ & 0.398$\pm$0.092 \\
    \textbf{Tucker} & 0.679$\pm$0.014 & 0.269$\pm$0.028 & $-$ & 0.426$\pm$0.052 & 0.683$\pm$0.040 & 0.259$\pm$0.079 & $-$ & 0.414$\pm$0.134 \\
    \textbf{GPST} & 0.661$\pm$0.014 & 0.305$\pm$ 0.023 & 1.005$\pm$0.021 & 0.449$\pm$0.040 & 0.681$\pm$0.043 & 0.265$\pm$0.087 & $\mathbf{1.035\pm0.061}$ & 0.412$\pm$0.112 \\
    \hline
    \end{tabular}}
    \label{tab:flare-results}
\end{table*}

% \begin{table}[htb]
%     \centering
%     \caption{Solar Flare intensity regression results on training (top) and testing (bottom) set across five different models under random train/test splitting. The metrics are mean-squared-error (MSE), R-squared, Coverage Probability ($\mbox{P}_{\mbox{cover}}$), True Skill Statistics (TSS). \textbf{GPST-high} achieves better testing set performance across all categories. The estimated $\widehat{\sigma}$ for \textbf{GP}, \textbf{GPST-low}, \textbf{GPST-high} are $0.667$, $0.668$ and $0.662$, respectively.}
%     \begin{tabular}{|c|c|c|c|c|}
%     \hline
%     Model & MSE & $\mathrm{R}^{2}$ & $\mbox{P}_{\mbox{cover}}$ & TSS\\\hline
%     \multirow{2}{*}{\textbf{GP}} &  $\mathbf{0.409}$ & $\mathbf{0.357}$ & $0.963$ & $\mathbf{0.487}$ \\\cline{2-5}
%                 &  $0.569$ & $0.159$ & $0.920$ & $0.364$  \\\hline
%     \multirow{2}{*}{\textbf{GPST-low}} & $0.439$ & $0.310$ & $\mathbf{0.966}$ & $0.451$ \\\cline{2-5}
%                       & $0.462$ & $0.260$ & $0.957$ & $0.441$ \\\hline
%     \multirow{2}{*}{\textbf{GPST-high}} & $0.432$ & $0.319$ & $0.959$ & $0.452$ \\\cline{2-5}
%      & $\mathbf{0.432}$ & $\mathbf{0.301}$ & $\mathbf{0.962}$ & $\mathbf{0.446}$  \\\hline
%     \multirow{2}{*}{\textbf{CP}} & $0.471$ & $0.260$ & $-$ & $0.408$ \\\cline{2-5}
%      & $0.457$ & $0.266$ & $-$ & $0.431$ \\\hline
%     \multirow{2}{*}{\textbf{Tucker}} & $0.470$ & $0.261$ & $-$ & $0.408$ \\\cline{2-5}
%      & $0.456$ & $0.266$ & $-$ & $0.431$ \\
%     \hline
%     \end{tabular}
%     \label{tab:flare-results}
% \end{table}

Tensor-GP (\textbf{GP}) shows worse generalizability on the testing data as compared to the other three methods. \textbf{GPST} has slightly better testing set performance compared to \textbf{CP} and \textbf{Tucker}, but is not statistically significantly better than \textbf{Tucker}. Similar to the simulation data, the flare data exhibits randomness of the location of flare predictive signals, making the tensor contraction a critical step for improving the Tensor-GP method.

\begin{figure}[htb]
    \centering
    \includegraphics[width=0.48\textwidth]{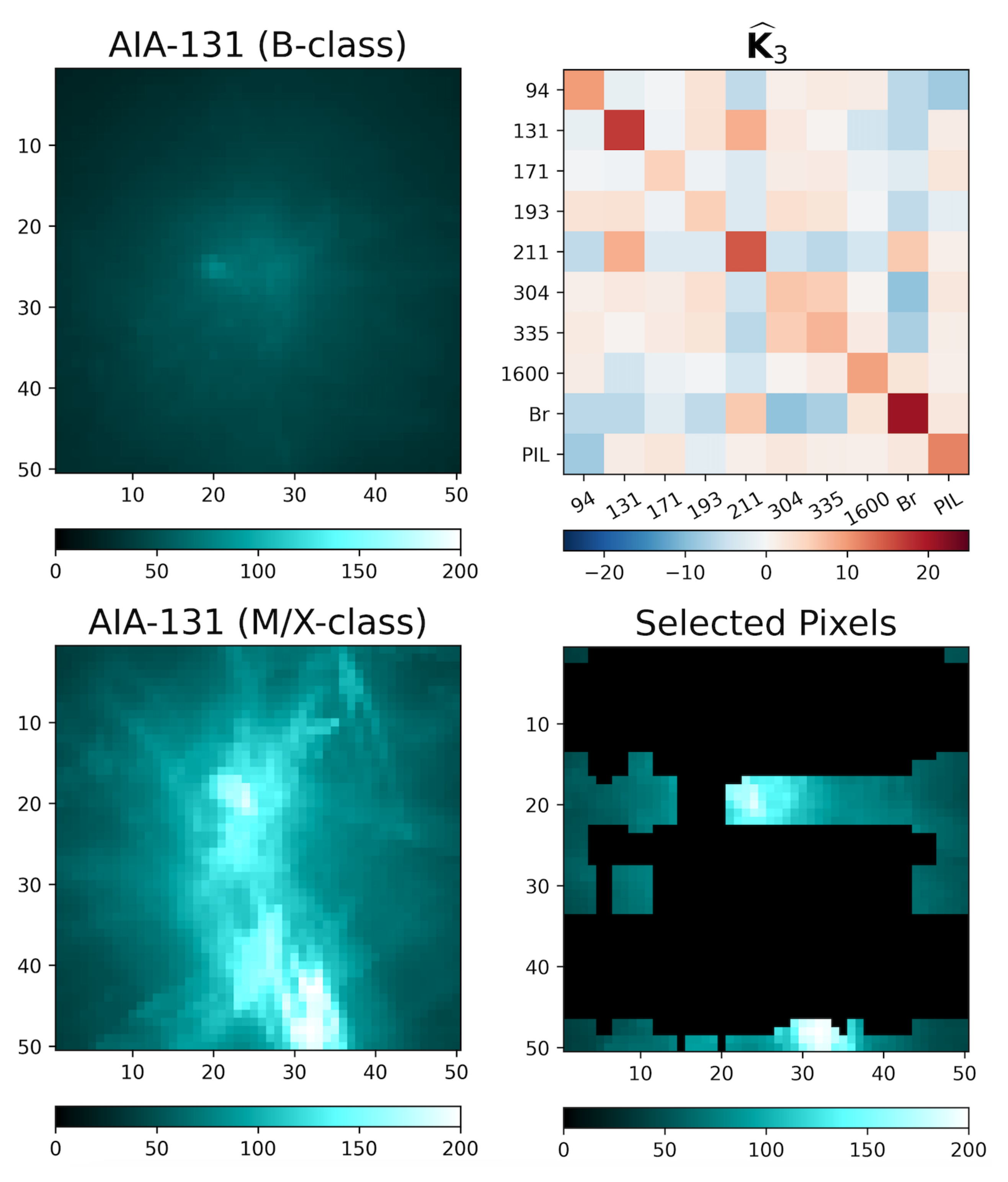}
    \caption{(Left column) The average AIA-131$\angstrom$ for all B-class flares and all M/X-class flares. (Right column) Estimated $\widehat{\mathbf{K}}_{3}$ in the multi-linear kernel that captures the channel-channel covariances (top). Pixels with at least one feature map with weight $>5\times 10^{-3}$ (bottom). We visualize the selected pixels with M/X class average AIA-131$\angstrom$ as the background. See full results in Appendix \ref{app:more-flare-results}.}
    \label{fig:flare-regression-results}
\end{figure}

In Figure \ref{fig:flare-regression-results}, we visualize the class-average AIA-$131\angstrom$ in the left column. There is a stark contrast between the two flare classes for this channel and many other channels as we show in Appendix \ref{app:more-flare-results}. A convenient output of our Tensor-GPST model is the direct estimation of channel covariances in the multi-linear kernel, and we visualize the estimated $\widehat{\mathbf{K}}_{3}$ in the Figure as well. The estimated $\widehat{\mathbf{K}}_{3}$ reveals the important channel pairs when defining the similarity of pairs of tensor data, and we formalize this channel pair importance notion in \eqref{eq:single-map-expvar} of Appendix \ref{app:more-flare-results}. To the best of our knowledge, our model is the first to consider the channel interactions for solar flare forecasting.

In the lower right panel of Figure \ref{fig:flare-regression-results}, we visualize the pixels that have at least one feature map with weight $>5\times 10^{-3}$. These pixels contribute significantly to building the latent tensor, and are thus being considered as the most relevant pixels for solar flare prediction. As one can see, the selected pixels are concentrated around the two brightest spots of the AIA-$131\angstrom$ for the M/X-class and also around the boundary. These pixels contain two most significant flare discriminating factors: 1) the brightest spots of the AIA images; 2) the span of the bright regions (as M/X flares still have large AIA image intensities near the boundary but not B flares).

% feature map with the highest $\%$ of explained variation (see definition in \eqref{eq:single-map-expvar}). The feature map captures the discriminating signals around the border, indicating that the learnt feature is collected around the perimeter of the flare eruptive region, which is confirmed by the class-average AIA-$131\angstrom$ shown in Figure \ref{fig:flare-regression-results}. The complete results are included in Appendix \ref{app:more-flare-results}.

\section{Conclusion}
In this paper, we propose a new methodology called \textbf{Tensor-GPST} for fitting Gaussian Process Regression (GPR) model on labelled multi-channel imaging data. We propose a tensor contraction operation to reduce the dimensionality of the tensor and also introduce anisotropic total variation penalty to the tensor contraction parameters to allow for interpretable feature extraction. We see improvements on the regression performances over the original Tensor-GP \cite{yu2018tensor} in both simulation and the solar flare forecasting task. The capability of the model in generating an interpretable low-dimensional tensor representation makes it ideal for many other scientific applications, such as predicting ADHD with Brain-Image \cite{li2018tucker} and studying the association of brain connectivity with human traits \cite{papadogeorgou2021soft}.

The current model has several limitations that can potentially lead to future research directions. First, we do not impose explicit identifiability constraints for the tensor contraction parameters and the multi-linear kernel parameters. This makes the optimization problem unconstrained thus enabling a simple gradient-based algorithm, but makes the parameters not fully identifiable. Second, the model has higher computational complexity as compared to the non-GP tensor regression models due to its GP formulation. A more efficient computational algorithm is needed to handle larger datasets.

Our code is available on GitHub at \url{https://github.com/husun0822/TensorGPST}.

\section*{Acknowledgement}
This material is based upon work supported by the National Science Foundation under Grant NSF-PHY 2027555 and NSF-DMS 113397. WBM is supported by NASA grant 80NSSC18K1208.

\newpage
\bibliography{references}
\bibliographystyle{icml2023}
%%%%%%%%%%%%%%%%%%%%%%%%%%%%%%%%%%%%%%%%%%%%%%%%%%%%%%%%%%%%%%%%%%%%%%%%%%%%%%%
%%%%%%%%%%%%%%%%%%%%%%%%%%%%%%%%%%%%%%%%%%%%%%%%%%%%%%%%%%%%%%%%%%%%%%%%%%%%%%%
% APPENDIX
%%%%%%%%%%%%%%%%%%%%%%%%%%%%%%%%%%%%%%%%%%%%%%%%%%%%%%%%%%%%%%%%%%%%%%%%%%%%%%%
%%%%%%%%%%%%%%%%%%%%%%%%%%%%%%%%%%%%%%%%%%%%%%%%%%%%%%%%%%%%%%%%%%%%%%%%%%%%%%%
\newpage
\appendix
\onecolumn
\section{Proof of Lemma \ref{thm:tv-norm-decomp}}\label{app:lemma-1}
The anisotropic total variation penalty can be simplified as follows, thanks to the rank-1 assumption on the feature map $\mathbf{W}_{s,t}$:
\begin{align}
    \|\mathbf{W}_{s,t}\|_{\text{TV}} & = \sum_{i=1}^{H-1}\sum_{j=1}^{W}  \left|\mathbf{W}_{s,t}(i+1,j)-\mathbf{W}_{s,t}(i,j)\right| + \sum_{i=1}^{H}\sum_{j=1}^{W-1}  \left|\mathbf{W}_{s,t}(i,j+1)-\mathbf{W}_{s,t}(i,j)\right| \\
    & = \sum_{i=1}^{H-1}\sum_{j=1}^{W} |\mathbf{A}(s,i+1)-\mathbf{A}(s,i)|\cdot|\mathbf{B}(t,j)| + \sum_{i=1}^{H}\sum_{j=1}^{W-1} |\mathbf{B}(t,j+1)-\mathbf{B}(t,j)|\cdot|\mathbf{A}(s,i)|.
\end{align}

As a result, the total variation penalty has an elegant multiplicative formulation:
\begin{align}
    \lambda \sum_{s=1}^{h}\sum_{t=1}^{w} \|\mathbf{W}_{s,t}\|_{\text{TV}} & = \left(\lambda\sum_{t=1}^{w}\sum_{j=1}^{W} \left|\mathbf{B}(t,j)\right|\right)\cdot\left(\sum_{s=1}^{h}\sum_{i=1}^{H-1}\left|\mathbf{A}(s,i+1)-\mathbf{A}(s,i)\right|\right) \nonumber \\
    & + \left(\lambda\sum_{t=1}^{w}\sum_{j=1}^{W-1} \left|\mathbf{B}(t,j+1)-\mathbf{B}(t,j)\right|\right)\cdot\left(\sum_{s=1}^{h}\sum_{i=1}^{H}\left|\mathbf{A}(s,i)\right|\right) \nonumber \\
    & = \lambda\cdot\|\mathbf{B}\|_{1}\cdot \|\nabla_{x}\mathbf{A}\|_{1} + \lambda\cdot\|\nabla_{x}\mathbf{B}\|_{1}\cdot\|\mathbf{A}\|_{1},
    \label{eq:fused-lasso}
\end{align}
where $\nabla_{x}$ is the horizontal (i.e. row) first-order derivative operator and $\|\cdot\|_{1}$ is the matrix $\ell_{1}$ norm. \eqref{eq:fused-lasso} turns out to be a fused-lasso type penalty \cite{tibshirani2005sparsity} with both a penalty on the sparsity of $\mathbf{A}$ and a penalty on the smoothness of each row of $\mathbf{A}$. This leads to a rank-1 feature map with row smoothness. Different from the fused-lasso penalty, we only introduce one tuning parameter $\lambda$ instead of $\lambda_{1},\lambda_{2}$ for the sparsity of smoothness of $\mathbf{A}$ separately. Instead, the tuning parameter for the sparsity of smoothness of $\mathbf{A}$ is re-weighted by the smoothness and sparsity of $\mathbf{B}$ and vice versa, according to \eqref{eq:fused-lasso}.

\section{Proximal Gradient Descent Algorithm for Tensor-GPST}\label{app:algorithm}

Given the factorization assumption for the tensor multi-linear kernel factors $\mathbf{K}_{1}, \mathbf{K}_{2}, \mathbf{K}_{3}$:
\begin{equation}\label{eq:Factorization_Assumption}
    \mathbf{K}_{1} = \mathbf{U}_{1}^{\top}\mathbf{U}_{1}, \mathbf{K}_{2} = \mathbf{U}_{2}^{\top}\mathbf{U}_{2}, \mathbf{K}_{3} = \mathbf{U}_{3}^{\top}\mathbf{U}_{3},
\end{equation}
where $\mathbf{U}_{1}\in \mathbb{R}^{r_{1}\times h}, \mathbf{U}_{2} \in \mathbb{R}^{r_{2}\times w}, \mathbf{U}_{3} \in \mathbb{R}^{r_{3}\times C}$. One can rewrite the penalized negative log-likelihood loss function in \eqref{eq:loss} as:
\begin{align}
    \min_{\mathbf{A}, \mathbf{B}, \mathbf{U}_{1:3}, \sigma} L = & \frac12 \log\left|\widetilde{\mathbf{U}}\widetilde{\mathbf{U}}^{\top} + \mathbf{D}_{\sigma}\right| + \frac12 \mathbf{y}^{\top} \left(\widetilde{\mathbf{U}}\widetilde{\mathbf{U}}^{\top} + \mathbf{D}_{\sigma}\right)^{-1}\mathbf{y} \nonumber \\
    & + \lambda\sum_{s=1}^{h}\sum_{t=1}^{w} \left\{\sum_{i=1}^{H-1}\sum_{j=1}^{W}  \left|\mathbf{W}_{s,t}(i+1,j)-\mathbf{W}_{s,t}(i,j)\right| + \sum_{i=1}^{H}\sum_{j=1}^{W-1}  \left|\mathbf{W}_{s,t}(i,j+1)-\mathbf{W}_{s,t}(i,j)\right|\right\} \label{eq:loss2} \\
    = & \ell(\mathbf{A}, \mathbf{B}, \mathbf{U}_{1:3}, \sigma) + \lambda \sum_{s=1}^{h}\sum_{t=1}^{w} \|\mathbf{W}_{s,t}\|_{\text{TV}}, \nonumber
\end{align}
recall that: 
$$
\mathbf{y} = [y_{1},y_{2},\ldots,y_{N}]^{\top}, \quad \widetilde{\mathcal{X}} = [\vect{\mathcal{X}_{1}}:\vect{\mathcal{X}_{2}}:\dots:\vect{\mathcal{X}_{N}}], \quad \widetilde{\mathbf{U}} = \widetilde{\mathcal{X}}^{\top}\left(\mathbf{I}_{C}\otimes \mathbf{B}\otimes \mathbf{A}\right)^{\top} \left(\mathbf{U}_{3}\otimes \mathbf{U}_{2} \otimes \mathbf{U}_{1}\right)^{\top}.
$$ 
We update the model parameters via a block coordinate descent scheme following the order of: $\mathbf{A} \rightarrow \mathbf{B} \rightarrow \left(\mathbf{U}_{1}, \mathbf{U}_{2}, \mathbf{U}_{3}\right) \rightarrow \sigma \rightarrow \mathbf{A} \rightarrow \mathbf{B} \rightarrow \dots$.

The derivation of the gradients of $\ell(\cdot)$ w.r.t. $\mathbf{A}, \mathbf{B}, \mathbf{U}_{1:3}, \sigma$ have been made trivial thanks to the factorization assumption \eqref{eq:Factorization_Assumption}. For the $(i,j)^{\rm{th}}$ element of $\mathbf{A}$, for instance, we have its partial derivative as:
$$
\frac{\partial \ell}{\partial \mathbf{A}}(i,j) = \mathbf{tr}\left[\left(\frac{\partial \ell}{\partial \widetilde{\mathbf{U}}}\right)^{\top}\left(\frac{\partial \widetilde{\mathbf{U}}}{\partial \mathbf{A}(i,j)}\right)\right], \quad \frac{\partial \widetilde{\mathbf{U}}}{\partial \mathbf{A}(i,j)} = \widetilde{\mathcal{X}}^{\top}\left(\mathbf{I}_{C}\otimes \mathbf{B}\otimes \mathbf{O}_{ij}\right)^{\top} \left(\mathbf{U}_{3}\otimes \mathbf{U}_{2} \otimes \mathbf{U}_{1}\right)^{\top}
$$
where $\mathbf{O}_{ij} \in \mathbb{R}^{h\times H}$ is a binary matrix with all entries being zero except the $(i,j)^{\rm{th}}$ entry being one. The derivative of $\ell$ w.r.t. $\widetilde{\mathbf{U}}$ has an explicit form \cite{yu2018tensor}:
\begin{equation*}
    \frac{\partial \ell}{\partial \widetilde{\mathbf{U}}} = \widetilde{\mathbf{U}}\left(\boldsymbol{\Sigma}^{-1} + \boldsymbol{\Sigma}^{-1}\widetilde{\mathbf{U}}^{\top}\mathbf{y}\mathbf{D}_{\sigma}^{-1}\mathbf{y}^{\top}\widetilde{\mathbf{U}}\boldsymbol{\Sigma}^{-1}\right) - \mathbf{y}\mathbf{D}_{\sigma}^{-1}\mathbf{y}^{\top}\widetilde{\mathbf{U}}\boldsymbol{\Sigma}^{-1}
\end{equation*}
where $\boldsymbol{\Sigma} = \mathbf{D}_{\sigma} + \widetilde{\mathbf{U}}^{\top}\widetilde{\mathbf{U}}$. The derivative of $\widetilde{\mathbf{U}}$ w.r.t. $\mathbf{A}, \mathbf{B}, \mathbf{U}_{1}, \mathbf{U}_{2}, \mathbf{U}_{3}$ can be readily derived by simply replacing each matrix parameter with a sparse binary matrix such as $\mathbf{O}_{ij}$ stated above. For example for $\mathbf{U}_{2}$, one has:
\begin{equation*}
    \frac{\partial \widetilde{\mathbf{U}}}{\partial \mathbf{U}_{2}(i,j)} = \widetilde{\mathcal{X}}^{\top}\left(\mathbf{I}_{C}\otimes \mathbf{B}\otimes \mathbf{A}\right)^{\top} \left(\mathbf{U}_{3}\otimes \mathbf{O}_{ij} \otimes \mathbf{U}_{1}\right)^{\top}
\end{equation*}
where $\mathbf{O}_{ij} \in \mathbb{R}^{r_{2}\times w}$ and is sparse except the $(i,j)^{\rm{th}}$ element being one. The gradients for $\mathbf{U}_{1}, \mathbf{U}_{2}, \mathbf{U}_{3}$ can be used for parameter update, and for $\mathbf{A}$ and $\mathbf{B}$, we consider updating them via proximal gradient descent. For $\mathbf{A}$ at the $(i+1)^{\rm{th}}$ iteration, for example, one applies the gradient descent first to get an estimator proposal: $\widehat{\mathbf{A}}^{(i+\frac12)} = \widehat{\mathbf{A}}^{(i)} - \eta_{i}\frac{\partial \ell}{\partial \mathbf{A}}$, and then solves the following optimization problem, which is commonly known as the proximal step:
\begin{equation}\label{eq:prox-A}
    \mathbf{prox}_{\text{TV}}(\widehat{\mathbf{A}}^{(k+\frac12)}) = \argmin_{\mathbf{A}\in\mathbb{R}^{h\times H}} \left\{\frac{1}{2\eta_{i}}\left\|\mathbf{A} - \widehat{\mathbf{A}}^{(i+\frac12)}\right\|_{\mathrm{F}}^{2} + \lambda \sum_{s=1}^{h}\sum_{t=1}^{w} \|\mathbf{W}_{s,t}\|_{\text{TV}}\right\}
\end{equation}
where $\eta_{i}$ is the learning rate of the $(i+1)^{\rm{th}}$ step.

Solving the proximal problem in \eqref{eq:prox-A} can be broken down into multiple parallel 1-D \textit{fused lasso signal approximation} problem. According to Proposition 1 of \citet{friedman2007pathwise}, solving \eqref{eq:prox-A} can be further broken down into first solving $h$ total variation de-noising problem \cite{rudin1992nonlinear}:
\begin{equation}\label{eq:prox-A-TV}
    \widetilde{\mathbf{A}}(s,:) \leftarrow \argmin_{\boldsymbol{\alpha} \in \mathbb{R}^{H}} \frac{1}{2\eta_{i}}\left\|\boldsymbol{\alpha} - \widehat{\mathbf{A}}^{(i+\frac12)}(s,:)\right\|_{\mathrm{F}}^{2} + \lambda \cdot \|\mathbf{B}\|_{1} \cdot \sum_{j=2}^{H}|\boldsymbol{\alpha}(j+1)-\boldsymbol{\alpha}(j)|, \quad s=1,2,\dots, h
\end{equation}
Then one can apply a soft-thresholding operator $\mathcal{S}_{\lambda\|\nabla_{x}\mathbf{B}\|_{1}}(\cdot)$, element-wisely, to $\widetilde{\mathbf{A}}\coloneqq \mathbf{prox}_{\text{TV}}(\widehat{\mathbf{A}}^{(k+\frac12)})$ to obtain the solution for \eqref{eq:prox-A}. The problem in \eqref{eq:prox-A-TV} can be efficiently solved via the python implementation in \texttt{prox-TV} based on a fast Newton's method \cite{jimenez2011fast,barbero2018modular}. Similar technique can be applied to update $\mathbf{B}$, and therefore the final optimization algorithm consists of both a gradient descent step and a fused-lasso proximal step. A more general theoretical discussion on the total variation penalty over 1-D signals can be found in \citet{tibshirani2014adaptive}.

The gradient of $\ell(\cdot)$ w.r.t. $\sigma^2$ can be easily derived as follows:

\begin{equation}\label{eq:grad-sigma}
\frac{\partial \ell}{\partial \sigma^{2}} = \mathbf{tr}\left[\left(\frac{\partial \ell}{\partial (\mathbf{K}+\mathbf{D}_{\sigma})^{-1}}\right)^{\top}\left(\frac{\partial (\mathbf{K}+\mathbf{D}_{\sigma})^{-1}}{\partial \sigma^{2}}\right)\right] = \mathbf{tr}\left[\frac12 \left(\mathbf{K}+\mathbf{D}_{\sigma}\right)^{-1} - \frac12 \left(\mathbf{K}+\mathbf{D}_{\sigma}\right)^{-2}\mathbf{y}\mathbf{y}^{\top}\right]
\end{equation}

Predictions on the unseen testing data with covariates $\mathbf{X}_{*}$, given the training data $(\mathbf{X},\mathbf{y})$, can be easily derived using the predictive distribution $\left(\mathbf{y}_{*}|\mathbf{X}_{*},\mathbf{X},\mathbf{y}\right) \sim \mathcal{N}\left(\boldsymbol{\mu}_{*},\boldsymbol{\Sigma}_{*}\right)$:
\begin{align*}
    \boldsymbol{\mu}_{*} & = \widehat{\mathcal{K}}\left(\mathbf{X}_{*}, \mathbf{X}\right)\left(\widehat{\mathcal{K}}\left(\mathbf{X}, \mathbf{X}\right)+\mathbf{D}_{\widehat{\sigma}}\right)^{-1}\mathbf{y} \\
    \boldsymbol{\Sigma}_{*} & = \widehat{\mathcal{K}}\left(\mathbf{X}_{*}, \mathbf{X}_{*}\right) + \mathbf{D}_{\widehat{\sigma}} - \widehat{\mathcal{K}}\left(\mathbf{X}_{*}, \mathbf{X}\right)\left(\widehat{\mathcal{K}}\left(\mathbf{X}, \mathbf{X}\right)+\mathbf{D}_{\widehat{\sigma}}\right)^{-1}\widehat{\mathcal{K}}\left(\mathbf{X}_{*}, \mathbf{X}\right)^{\top}
\end{align*}
where $\widehat{\mathcal{K}}(\cdot,\cdot)$ is the kernel function in \eqref{eq:MLGP-X-Kernel} but evaluated at the estimated model parameters, and $\widehat{\mathcal{K}}(\mathbf{X}_{*},\mathbf{X})$ simply denotes the covariances between the unseen data $\mathbf{X}_{*}$ and the training data $\mathbf{X}$, and the other notations follow.

%%%%%%%%%%%%%%%%%%%%%%%%%%%%%%%%%%%%%%%%%%%%%%%%%%%%%%%%%%%%%%%%%%%%%%%%%%%%%%%
%%%%%%%%%%%%%%%%%%%%%%%%%%%%%%%%%%%%%%%%%%%%%%%%%%%%%%%%%%%%%%%%%%%%%%%%%%%%%%%

\section{Proof of Theorem \ref{thm:conv-rate}}\label{app:proof-conv-rate}
The proof of Theorem \ref{thm:conv-rate} is largely based on the convergence results of proximal gradient descent but with additional consideration on the alternating descent scheme. We denote an arbitrary collection of model parameters as $\boldsymbol{\theta} \coloneqq \left(\mathbf{A},\mathbf{B},\mathbf{U}_{1:3},\sigma\right)$. Since we update the four blocks of parameters cyclically in Algorithm \ref{algorithm}, within each iteration, we further denote the intermediate parameter updates as $\widehat{\boldsymbol{\theta}}^{(k)} \xrightarrow{\mbox{update }\mathbf{A}} \widehat{\boldsymbol{\theta}}^{(k+\frac14)} \xrightarrow{\mbox{update }\mathbf{B}} \widehat{\boldsymbol{\theta}}^{(k+\frac12)} \xrightarrow{\mbox{update } \mathbf{U}_{1:3}} \widehat{\boldsymbol{\theta}}^{(k+\frac34)} \xrightarrow{\mbox{update } \sigma} \widehat{\boldsymbol{\theta}}^{(k+1)}$. In the remainder of the proof, we will use $\mathbf{U}$ to denote $\mathbf{U}_{1:3}$ for notational simplicity.

In order to show the upper bound of the difference of the loss function after $K$ iterations with its global minimum $L(\boldsymbol{\theta}^{*})$, we first show Lemma \ref{thm:lemma-1-app}:

\begin{lemma}\label{thm:lemma-1-app}
Given the alternating proximal gradient descent algorithm in Algorithm \ref{algorithm} and the assumptions made in Theorem \ref{thm:conv-rate}, one can bound $L(\widehat{\boldsymbol{\theta}}^{(k+\frac{v}{4})}), v\in \{1,2,3,4\}$ as:
\begin{align}
    L(\widehat{\boldsymbol{\theta}}^{(k+\frac{1}{4})}) & \le L(\mathbf{A}^{*}, \widehat{\mathbf{B}}^{(k)}, \widehat{\mathbf{U}}^{(k)}, \widehat{\sigma}^{(k)}) + \frac{1}{2\alpha} \left\{\left\|\widehat{\mathbf{A}}^{(k)} - \mathbf{A}^{*}\right\|^{2} - \left\|\widehat{\mathbf{A}}^{(k+1)} - \mathbf{A}^{*}\right\|^{2}\right\} \label{eq:A-update-bound} \\
    L(\widehat{\boldsymbol{\theta}}^{(k+\frac{1}{2})}) & \le L(\widehat{\mathbf{A}}^{(k+1)}, \mathbf{B}^{(*)}, \widehat{\mathbf{U}}^{(k)}, \widehat{\sigma}^{(k)}) + \frac{1}{2\alpha} \left\{\left\|\widehat{\mathbf{B}}^{(k)} - \mathbf{B}^{*}\right\|^{2} - \left\|\widehat{\mathbf{B}}^{(k+1)} - \mathbf{B}^{*}\right\|^{2}\right\} \\
    L(\widehat{\boldsymbol{\theta}}^{(k+\frac{3}{4})}) & \le L(\widehat{\mathbf{A}}^{(k+1)}, \widehat{\mathbf{B}}^{(k+1)}, \mathbf{U}^{*}, \widehat{\sigma}^{(k)}) + \frac{1}{2\alpha} \left\{\left\|\widehat{\mathbf{U}}^{(k)} - \mathbf{U}^{*}\right\|^{2} - \left\|\widehat{\mathbf{U}}^{(k+1)} - \mathbf{U}^{*}\right\|^{2}\right\} \\
    L(\widehat{\boldsymbol{\theta}}^{(k+1)}) & \le L(\widehat{\mathbf{A}}^{(k+1)}, \widehat{\mathbf{B}}^{(k+1)}, \widehat{\mathbf{U}}^{(k+1)}, \sigma^{*}) + \frac{1}{2\alpha} \left\{\left\|\widehat{\sigma}^{(k)} - \sigma^{*}\right\|^{2} - \left\|\widehat{\sigma}^{(k+1)} - \sigma^{*}\right\|^{2}\right\}, 
\end{align}
where $\|\cdot\|$ is the matrix Frobenious norm, $\alpha$ is a constant learning rate with $\alpha \le 1/\max\{M_{\mathbf{A}}, M_{\mathbf{B}}, M_{\mathbf{U}}, M_{\sigma}\}$ and $M_{\mathbf{A}}, M_{\mathbf{B}}, M_{\mathbf{U}}, M_{\sigma}$ are the Lipchitz constant for $\mathbf{A},\mathbf{B},\mathbf{U},\sigma$ for the gradient of $\ell(\cdot)$, i.e. the negative log-likelihood defined in \eqref{eq:Negative-LL}.
\end{lemma}

\begin{proof}
It suffices to prove \eqref{eq:A-update-bound}, and the rest of the bounds follow the same technique.

First, given that the gradient of $\ell(\cdot)$ w.r.t. to $\mathbf{A}$ is Lipschitz continuous with constant $M_{\mathbf{A}}$, i.e. $\|\nabla_{\mathbf{A}}\ell(\mathbf{A}_{1}) - \nabla_{\mathbf{A}}\ell(\mathbf{A}_{2})\| \le M_{\mathbf{A}}\|\mathbf{A}_{1}- \mathbf{A}_{2}\|, \forall \mathbf{A}_{1}, \mathbf{A}_{2}$. Since the other parameters also share the same property but have different Lipschitz constant $M_{\mathbf{B}}, M_{\mathbf{U}}, M_{\sigma}$, we use $M \coloneqq \max\{M_{\mathbf{A}},M_{\mathbf{B}}, M_{\mathbf{U}}, M_{\sigma}\}$ as the Lipschitz constant for all parameters. Given the Lipschitz continuity of the derivative, one has:
\begin{equation}\label{eq:theta-ub-1}
\ell(\widehat{\boldsymbol{\theta}}^{(k+\frac14)}) \le \ell(\widehat{\boldsymbol{\theta}}^{(k)}) + \left<\nabla_{\mathbf{A}}\ell(\widehat{\boldsymbol{\theta}}^{(k)}), \widehat{\mathbf{A}}^{(k+1)}-\widehat{\mathbf{A}}^{(k)}\right> + \frac{M}{2} \|\widehat{\mathbf{A}}^{(k+1)}-\widehat{\mathbf{A}}^{(k)}\|^{2}
\end{equation}
which is a direct result from the following inequality for any function $\ell(\cdot)$ with $M$-Lipschitz continuous derivative:
$$
\ell(y) \le \ell(x) + \left<\nabla_{x}\ell(x), y-x\right> + \frac{M}{2} \|y-x\|^{2}
$$
Additionally, since $\ell(\cdot)$ is assumed as block-wise convex, one has a natural upper bound of $\ell(\widehat{\boldsymbol{\theta}}^{(k)})$ based on convexity:
\begin{equation}\label{eq:theta-ub-2}
    \ell(\widehat{\boldsymbol{\theta}}^{(k)}) \le \ell(\mathbf{A}^{*}, \widehat{\mathbf{B}}^{(k)}, \widehat{\mathbf{U}}^{(k)}, \widehat{\sigma}^{(k)}) - \left<\nabla_{\mathbf{A}}\ell(\widehat{\boldsymbol{\theta}}^{(k)}), \mathbf{A}^{*}-\widehat{\mathbf{A}}^{(k)}\right>
\end{equation}

Combining \eqref{eq:theta-ub-1} and \eqref{eq:theta-ub-2}, one obtains:
\begin{equation}\label{eq:ll-bound}
\ell(\widehat{\boldsymbol{\theta}}^{(k+\frac14)}) \le \ell(\mathbf{A}^{*}, \widehat{\mathbf{B}}^{(k)}, \widehat{\mathbf{U}}^{(k)}, \widehat{\sigma}^{(k)}) + \left<\nabla_{\mathbf{A}}\ell(\widehat{\boldsymbol{\theta}}^{(k)}), \widehat{\mathbf{A}}^{(k+1)}-\mathbf{A}^{*}\right> + \frac{M}{2} \|\widehat{\mathbf{A}}^{(k+1)}-\widehat{\mathbf{A}}^{(k)}\|^{2}
\end{equation}

Also, since $\widehat{\mathbf{A}}^{(k+1)}$ is obtained via a proximal step:
\begin{equation*}
\widehat{\mathbf{A}}^{(k+1)} = \argmin_{\mathbf{A}} \frac{1}{2\alpha} \left\|\mathbf{A} - \left(\widehat{\mathbf{A}}^{(k)} - \alpha \nabla_{\mathbf{A}}\ell(\widehat{\boldsymbol{\theta}}^{(k)})\right)\right\|^{2} + \lambda \mathrm{R}\left(\mathbf{A}, \widehat{\mathbf{B}}^{(k)}\right)
\end{equation*}
$\widehat{\mathbf{A}}^{(k+1)}$ should satisfy the following subgradient condition:
\begin{equation}
    \mathrm{G}_{\alpha}(\widehat{\boldsymbol{\theta}}^{(k)}) - \nabla_{\mathbf{A}}\ell(\widehat{\boldsymbol{\theta}}^{(k)}) \in \lambda\cdot \partial_{A}\mathrm{R}\left(\widehat{\mathbf{A}}^{(k+1)}, \widehat{\mathbf{B}}^{(k)}\right)
\end{equation}
where $\mathrm{G}_{\alpha}(\widehat{\boldsymbol{\theta}}^{(k)}) \coloneqq -\frac{1}{\alpha}\left(\widehat{\mathbf{A}}^{(k+1)} - \widehat{\mathbf{A}}^{(k)}\right)$ is the proximal gradient. Using the definition of subgradient, one can achieve a trivial inequality as follows:
\begin{equation}\label{eq:subgrad-bound}
    \lambda\mathrm{R}\left(\widehat{\mathbf{A}}^{(k+1)}, \widehat{\mathbf{B}}^{(k)}\right) + \left<\mathrm{G}_{\alpha}(\widehat{\boldsymbol{\theta}}^{(k)}) - \nabla_{\mathbf{A}}\ell(\widehat{\boldsymbol{\theta}}^{(k)}), \mathbf{A}^{*}-\widehat{\mathbf{A}}^{(k+1)}\right> \le \lambda\mathrm{R}\left(\mathbf{A}^{*}, \widehat{\mathbf{B}}^{(k)}\right)
\end{equation}

Combining \eqref{eq:ll-bound} and \eqref{eq:subgrad-bound}, we have:
\begin{align*}
L(\widehat{\boldsymbol{\theta}}^{(k+\frac14)}) & = \ell(\widehat{\boldsymbol{\theta}}^{(k+\frac14)}) + \lambda \mathrm{R}\left(\widehat{\mathbf{A}}^{(k+1)}, \widehat{\mathbf{B}}^{(k)}\right) \\
& \le     L(\mathbf{A}^{*}, \widehat{\mathbf{B}}^{(k)}, \widehat{\mathbf{U}}^{(k)}, \widehat{\sigma}^{(k)}) + \left<\mathrm{G}_{\alpha}(\widehat{\boldsymbol{\theta}}^{(k)}), \widehat{\mathbf{A}}^{(k+1)}-\mathbf{A}^{*}\right> + \frac{M}{2} \|\widehat{\mathbf{A}}^{(k+1)}-\widehat{\mathbf{A}}^{(k)}\|^{2} \\
& \le L(\mathbf{A}^{*}, \widehat{\mathbf{B}}^{(k)}, \widehat{\mathbf{U}}^{(k)}, \widehat{\sigma}^{(k)}) + \left<\mathrm{G}_{\alpha}(\widehat{\boldsymbol{\theta}}^{(k)}), \widehat{\mathbf{A}}^{(k)}-\alpha \mathrm{G}_{\alpha}(\widehat{\boldsymbol{\theta}}^{(k)})-\mathbf{A}^{*}\right>  +\frac{1}{2\alpha}\left\|\alpha \mathrm{G}_{\alpha}(\widehat{\boldsymbol{\theta}}^{(k)})\right\|^{2} \\
& = L(\mathbf{A}^{*}, \widehat{\mathbf{B}}^{(k)}, \widehat{\mathbf{U}}^{(k)}, \widehat{\sigma}^{(k)}) + \frac{1}{2\alpha} \left\{\left\|\widehat{\mathbf{A}}^{(k)} - \mathbf{A}^{*}\right\|^{2} - \left\|\widehat{\mathbf{A}}^{(k+1)} - \mathbf{A}^{*}\right\|^{2}\right\}
\end{align*}
which completes the proof.
\end{proof}

In the classical proximal gradient descent context, where one updates a single parameter iteratively, the bound in Lemma \ref{thm:lemma-1-app} leads to a convergence rate of the algorithm at $\mathcal{O}(1/K)$, after one adds up all the inequalities from iteration $1$ to $K$. The key difference is that, on the right hand side of the inequality \eqref{eq:A-update-bound}, the loss function is evaluated at the global minima of $\mathbf{A}$ and the value of $\mathbf{B},\mathbf{U},\sigma$ \textit{at the $k^{\rm{th}}$ iteration}. We need to quantify the difference between $L(\mathbf{A}^{*}, \widehat{\mathbf{B}}^{(k)}, \widehat{\mathbf{U}}^{(k)}, \widehat{\sigma}^{(k)})$ and $L(\boldsymbol{\theta}^{*})$ to reach the final error bound result and this difference is given in Lemma \ref{thm:loss-additional-bound} below.

\begin{lemma}\label{thm:loss-additional-bound}
The difference of $L(\mathbf{A}^{*}, \widehat{\mathbf{B}}^{(k)}, \widehat{\mathbf{U}}^{(k)}, \widehat{\sigma}^{(k)})$ and $L(\boldsymbol{\theta}^{*}) \coloneqq L(\mathbf{A}^{*}, \mathbf{B}^{*}, \mathbf{U}^{*}, \sigma^{*})$ can be fully characterized as:
\begin{align}
L(\mathbf{A}^{*}, \widehat{\mathbf{B}}^{(k)}, \widehat{\mathbf{U}}^{(k)}, \widehat{\sigma}^{(k)}) - L(\boldsymbol{\theta}^{*}) & \le \frac{M}{2}\left\{\|\widehat{\mathbf{B}}^{(k)}-\mathbf{B}^{*}\| + \|\widehat{\mathbf{U}}^{(k)}-\mathbf{U}^{*}\| + \|\widehat{\sigma}^{(k)}-\sigma^{*}\|\right\}^{2} \label{eq:ALT-Bound} \\
& + \|\nabla_{\mathbf{B}}\ell(\boldsymbol{\theta}^{*})\|\cdot \|\widehat{\mathbf{B}}^{(k)}-\mathbf{B}^{*}\| + \lambda \mathrm{R}\left(\mathbf{A}^{*},\widehat{\mathbf{B}}^{(k)}-\mathbf{B}^{*}\right) \label{eq:TV-Bound}
\end{align}
where \eqref{eq:ALT-Bound} is the additional loss incurred by using the iterative value of the other parameters instead of the global optimum (called the ALT-gap), and \eqref{eq:TV-Bound} is the additional loss incurred by using the total variation penalty with the $k^{\rm{th}}$ iterative value of $\mathbf{B}$ (called the TV-gap).
\end{lemma}

\begin{proof}
We start the derivation with the following trivial decomposition:
\begin{align}
L(\mathbf{A}^{*}, \widehat{\mathbf{B}}^{(k)}, \widehat{\mathbf{U}}^{(k)}, \widehat{\sigma}^{(k)}) - L(\boldsymbol{\theta}^{*}) & = L(\mathbf{A}^{*}, \widehat{\mathbf{B}}^{(k)}, \widehat{\mathbf{U}}^{(k)}, \widehat{\sigma}^{(k)}) -  L(\mathbf{A}^{*}, \mathbf{B}^{*}, \widehat{\mathbf{U}}^{(k)}, \widehat{\sigma}^{(k)})  \label{eq:lemmaC2-bound-1} \\
& + L(\mathbf{A}^{*}, \mathbf{B}^{*}, \widehat{\mathbf{U}}^{(k)}, \widehat{\sigma}^{(k)}) - L(\mathbf{A}^{*}, \mathbf{B}^{*}, \mathbf{U}^{*}, \widehat{\sigma}^{(k)}) \label{eq:lemmaC2-bound-2}  \\
& + L(\mathbf{A}^{*}, \mathbf{B}^{*}, \mathbf{U}^{*}, \widehat{\sigma}^{(k)}) - L(\boldsymbol{\theta}^{*}) \label{eq:lemmaC2-bound-3}
\end{align}

We need to bound \eqref{eq:lemmaC2-bound-1}, \eqref{eq:lemmaC2-bound-2} and \eqref{eq:lemmaC2-bound-3} separately. For \eqref{eq:lemmaC2-bound-1}, we have:
\begin{align}
& L(\mathbf{A}^{*}, \widehat{\mathbf{B}}^{(k)}, \widehat{\mathbf{U}}^{(k)}, \widehat{\sigma}^{(k)}) -  L(\mathbf{A}^{*}, \mathbf{B}^{*}, \widehat{\mathbf{U}}^{(k)}, \widehat{\sigma}^{(k)}) \nonumber\\
& \le \ell(\mathbf{A}^{*}, \widehat{\mathbf{B}}^{(k)}, \widehat{\mathbf{U}}^{(k)}, \widehat{\sigma}^{(k)}) - \ell(\mathbf{A}^{*}, \mathbf{B}^{*}, \widehat{\mathbf{U}}^{(k)}, \widehat{\sigma}^{(k)}) + \lambda\mathrm{R}\left(\mathbf{A}^{*}, \widehat{\mathbf{B}}^{(k)}-\mathbf{B}^{*}\right) \nonumber \\
& \le \left<\nabla_{\mathbf{B}}\ell(\mathbf{A}^{*}, \mathbf{B}^{*}, \widehat{\mathbf{U}}^{(k)}, \widehat{\sigma}^{(k)}), \widehat{\mathbf{B}}^{(k)}-\mathbf{B}^{*}\right> + \frac{M}{2} \|\widehat{\mathbf{B}}^{(k)}-\mathbf{B}^{*}\|^{2} + \lambda\mathrm{R}\left(\mathbf{A}^{*}, \widehat{\mathbf{B}}^{(k)}-\mathbf{B}^{*}\right) \nonumber \\
& \le \|\widehat{\mathbf{B}}^{(k)}-\mathbf{B}^{*}\|\cdot \left(\|\nabla_{\mathbf{B}}\ell(\boldsymbol{\theta}^{*})\| + M\|\widehat{\mathbf{U}}^{(k)}-\mathbf{U}^{*}\| + M\|\widehat{\sigma}^{(k)}-\sigma^{*}\|\right) + \frac{M}{2} \|\widehat{\mathbf{B}}^{(k)}-\mathbf{B}^{*}\|^{2} + \lambda\mathrm{R}\left(\mathbf{A}^{*}, \widehat{\mathbf{B}}^{(k)}-\mathbf{B}^{*}\right)
\end{align}
where the last line follows from the Cauchy-Schwartz inequality followed by the Lipschitz continuity of the gradient and the triangle inequality of the Frobenius norm.

As for \eqref{eq:lemmaC2-bound-2}, similar technique follows and lead to:
\begin{align*}
L(\mathbf{A}^{*}, \mathbf{B}^{*}, \widehat{\mathbf{U}}^{(k)}, \widehat{\sigma}^{(k)}) - L(\mathbf{A}^{*}, \mathbf{B}^{*}, \mathbf{U}^{*}, \widehat{\sigma}^{(k)}) & = \ell(\mathbf{A}^{*}, \mathbf{B}^{*}, \widehat{\mathbf{U}}^{(k)}, \widehat{\sigma}^{(k)}) - \ell(\mathbf{A}^{*}, \mathbf{B}^{*}, \mathbf{U}^{*}, \widehat{\sigma}^{(k)}) \\
& \le \left<\nabla_{\mathbf{U}}\ell(\mathbf{A}^{*}, \mathbf{B}^{*}, \mathbf{U}^{*}, \widehat{\sigma}^{(k)}), \widehat{\mathbf{U}}^{(k)}-\mathbf{U}^{*}\right> + \frac{M}{2}\|\widehat{\mathbf{U}}^{(k)}-\mathbf{U}^{*}\|^{2} \\
& \le M\|\widehat{\mathbf{U}}^{(k)}-\mathbf{U}^{*}\|\cdot \|\widehat{\sigma}^{(k)}-\sigma^{*}\| + \frac{M}{2}\|\widehat{\mathbf{U}}^{(k)}-\mathbf{U}^{*}\|^{2}
\end{align*}
where the last line uses the fact that at the global optimum, we have $\nabla_{\mathbf{U}}\ell(\boldsymbol{\theta}^{*}) = 0$.

Similarly for \eqref{eq:lemmaC2-bound-3}, one has:
\begin{equation*}
L(\mathbf{A}^{*}, \mathbf{B}^{*}, \mathbf{U}^{*}, \widehat{\sigma}^{(k)}) - L(\boldsymbol{\theta}^{*}) \le \left<\nabla_{\sigma}\ell(\boldsymbol{\theta}^{*}), \widehat{\sigma}^{(k)} - \sigma^{*}\right> + \frac{M}{2}\|\widehat{\sigma}^{(k)} - \sigma^{*}\|^{2}    
\end{equation*}

Combining the three individual upper bounds together yields the result and thereby completes the proof.
\end{proof}

Similar results in Lemma \ref{thm:loss-additional-bound} can be easily derived for $\mathbf{B}^{*}, \mathbf{U}^{*}, \sigma^{*}$. With these theoretical results, we are now ready to prove Theorem \ref{thm:conv-rate}:

\begin{proof}
Combining the results in Lemma \ref{thm:lemma-1-app} and \ref{thm:loss-additional-bound}, we have the following upper bound for $L(\widehat{\boldsymbol{\theta}}^{(k+\frac14)}) - L(\boldsymbol{\theta}^{*})$:
\begin{align*}
L(\widehat{\boldsymbol{\theta}}^{(k+\frac14)}) - L(\boldsymbol{\theta}^{*}) & \le    \frac{M}{2}\left\{\|\widehat{\mathbf{B}}^{(k)}-\mathbf{B}^{*}\| + \|\widehat{\mathbf{U}}^{(k)}-\mathbf{U}^{*}\| + \|\widehat{\sigma}^{(k)}-\sigma^{*}\|\right\}^{2} \\
& + \|\nabla_{\mathbf{B}}\ell(\boldsymbol{\theta}^{*})\|\cdot \|\widehat{\mathbf{B}}^{(k)}-\mathbf{B}^{*}\| + \lambda \mathrm{R}\left(\mathbf{A}^{*},\widehat{\mathbf{B}}^{(k)}-\mathbf{B}^{*}\right) \\
& + \frac{1}{2\alpha} \left\{\left\|\widehat{\mathbf{A}}^{(k)} - \mathbf{A}^{*}\right\|^{2} - \left\|\widehat{\mathbf{A}}^{(k+1)} - \mathbf{A}^{*}\right\|^{2}\right\} 
\end{align*}

If $\mathbf{B}, \mathbf{U}, \sigma$ is fixed at $\mathbf{B}^{*}, \mathbf{U}^{*}, \sigma^{*}$ and one only updates $\mathbf{A}$ via the proximal gradient descent, the error bound here vanishes to its last term only and can be further reduced if one adds up the inequality from iteration $1$ to $K$, leading to the classical proximal gradient descent convergence rate result. Similar bounds for $L(\widehat{\boldsymbol{\theta}}^{(k+\frac12)}), L(\widehat{\boldsymbol{\theta}}^{(k+\frac34)}), L(\widehat{\boldsymbol{\theta}}^{(k+1)})$ can be derived and we aggregate the results together as follows:
\begin{align}\label{eq:loss-final-bound}
\sum_{k=0}^{K}\sum_{v=1}^{4} \left(L(\widehat{\boldsymbol{\theta}}^{(k+\frac{v}{4})}) - L(\boldsymbol{\theta}^{*})\right) &  \le \frac{1}{2\alpha} \left\{\|\widehat{\boldsymbol{\theta}}^{(0)}-\boldsymbol{\theta}^{*}\|^{2} - \|\widehat{\boldsymbol{\theta}}^{(K)}-\boldsymbol{\theta}^{*}\|^{2}\right\} \nonumber \\
& + \sum_{k=0}^{K} h_{\lambda}(\widehat{\mathbf{A}}^{(k+1)}-\mathbf{A}^{*}, \widehat{\mathbf{B}}^{(k+1)}-\mathbf{B}^{*}, \widehat{\mathbf{B}}^{(k)}-\mathbf{B}^{*}) \nonumber\\
& + \frac{1}{2\alpha} \sum_{k=0}^{K} \tau\left(\widehat{\boldsymbol{\theta}}^{(k+1)}, \widehat{\boldsymbol{\theta}}^{(k)}, \boldsymbol{\theta}^{*}\right) 
\end{align}
where $h_{\lambda}(\widehat{\mathbf{A}}^{(k+1)}-\mathbf{A}^{*}, \widehat{\mathbf{B}}^{(k+1)}-\mathbf{B}^{*}, \widehat{\mathbf{B}}^{(k)}-\mathbf{B}^{*})$ is the extra gap from the optimal loss created by the existence of the total variation penalty in the loss function (thus we call it the TV-gap) and is defined as:
\begin{align}\label{eq:TV-gap-form}
    h_{\lambda}(\widehat{\mathbf{A}}^{(k+1)}-\mathbf{A}^{*}, \widehat{\mathbf{B}}^{(k+1)}-\mathbf{B}^{*}, \widehat{\mathbf{B}}^{(k)}-\mathbf{B}^{*}) & \coloneqq 3\|\nabla_{\mathbf{A}}\ell(\boldsymbol{\theta}^{*})\|\cdot \|\widehat{\mathbf{A}}^{(k+1)}-\mathbf{A}^{*}\| + 2\|\nabla_{\mathbf{B}}\ell(\boldsymbol{\theta}^{*})\|\cdot\|\widehat{\mathbf{B}}^{(k+1)}-\mathbf{B}^{*}\| \nonumber \\
    & + \|\nabla_{\mathbf{B}}\ell(\boldsymbol{\theta}^{*})\|\cdot \|\widehat{\mathbf{B}}^{(k)}-\mathbf{B}^{*}\| + \lambda\mathrm{R}\left(\mathbf{A}^{*},\widehat{\mathbf{B}}^{(k)}-\mathbf{B}^{*}\right) \nonumber \\
    & + 3\lambda\mathrm{R}\left(\widehat{\mathbf{A}}^{(k+1)}-\mathbf{A}^{*},\mathbf{B}^{*}\right) + 2\lambda\mathrm{R}\left(\mathbf{A}^{*},\widehat{\mathbf{B}}^{(k+1)}-\mathbf{B}^{*}\right) \nonumber \\
    & + 2\lambda\mathrm{R}\left(\widehat{\mathbf{A}}^{(k+1)}-\mathbf{A}^{*},\widehat{\mathbf{B}}^{(k+1)}-\mathbf{B}^{*}\right)
\end{align}
where $\mathrm{R}(\mathbf{A}, \mathbf{B})$ is the total variation penalty defined in Lemma \ref{thm:tv-norm-decomp}. $\tau(\widehat{\boldsymbol{\theta}}^{(k+1)}, \widehat{\boldsymbol{\theta}}^{(k)}, \boldsymbol{\theta}^{*})$ is the extra gap from the optimal loss created by the usage of iterative value of the parameters during the alternating proximal gradient descent (thus we call it the ALT-gap) and is defined as:
\begin{align}
\tau\left(\widehat{\boldsymbol{\theta}}^{(k+1)}, \widehat{\boldsymbol{\theta}}^{(k)}, \boldsymbol{\theta}^{*}\right) & \coloneqq \left[\|\widehat{\mathbf{B}}^{(k)}-\mathbf{B}^{*}\| + \|\widehat{\mathbf{U}}^{(k)}-\mathbf{U}^{*}\| + \|\widehat{\sigma}^{(k)}-\sigma^{*}\|\right]^{2} \nonumber\\
& + \left[\|\widehat{\mathbf{A}}^{(k+1)}-\mathbf{A}^{*}\| + \|\widehat{\mathbf{U}}^{(k)}-\mathbf{U}^{*}\| + \|\widehat{\sigma}^{(k)}-\sigma^{*}\|\right]^{2}  \nonumber\\ 
&  + \left[\|\widehat{\mathbf{A}}^{(k+1)}-\mathbf{A}^{*}\| + \|\widehat{\mathbf{B}}^{(k+1)}-\mathbf{B}^{*}\| + \|\widehat{\sigma}^{(k)}-\sigma^{*}\|\right]^{2} \nonumber\\
& + \left[\|\widehat{\mathbf{A}}^{(k+1)}-\mathbf{A}^{*}\| + \|\widehat{\mathbf{B}}^{(k+1)}-\mathbf{B}^{*}\| + \|\widehat{\mathbf{U}}^{(k+1)}-\mathbf{U}^{*}\|\right]^{2} \label{eq:ALT-gap-form}
\end{align}

The final result can be derived from \eqref{eq:loss-final-bound} by lower bounding the left hand side:
$$
\sum_{k=0}^{K}\sum_{v=1}^{4} \left(L(\widehat{\boldsymbol{\theta}}^{(k+\frac{v}{4})}) - L(\boldsymbol{\theta}^{*})\right) \ge 4(K+1) \left(L(\widehat{\boldsymbol{\theta}}^{(K+1)}) - L(\boldsymbol{\theta}^{*})\right)
$$
which is evident given that each step is a descent step.
\end{proof}

\begin{figure}[htb]
    \centering
    \includegraphics[width=0.98\textwidth]{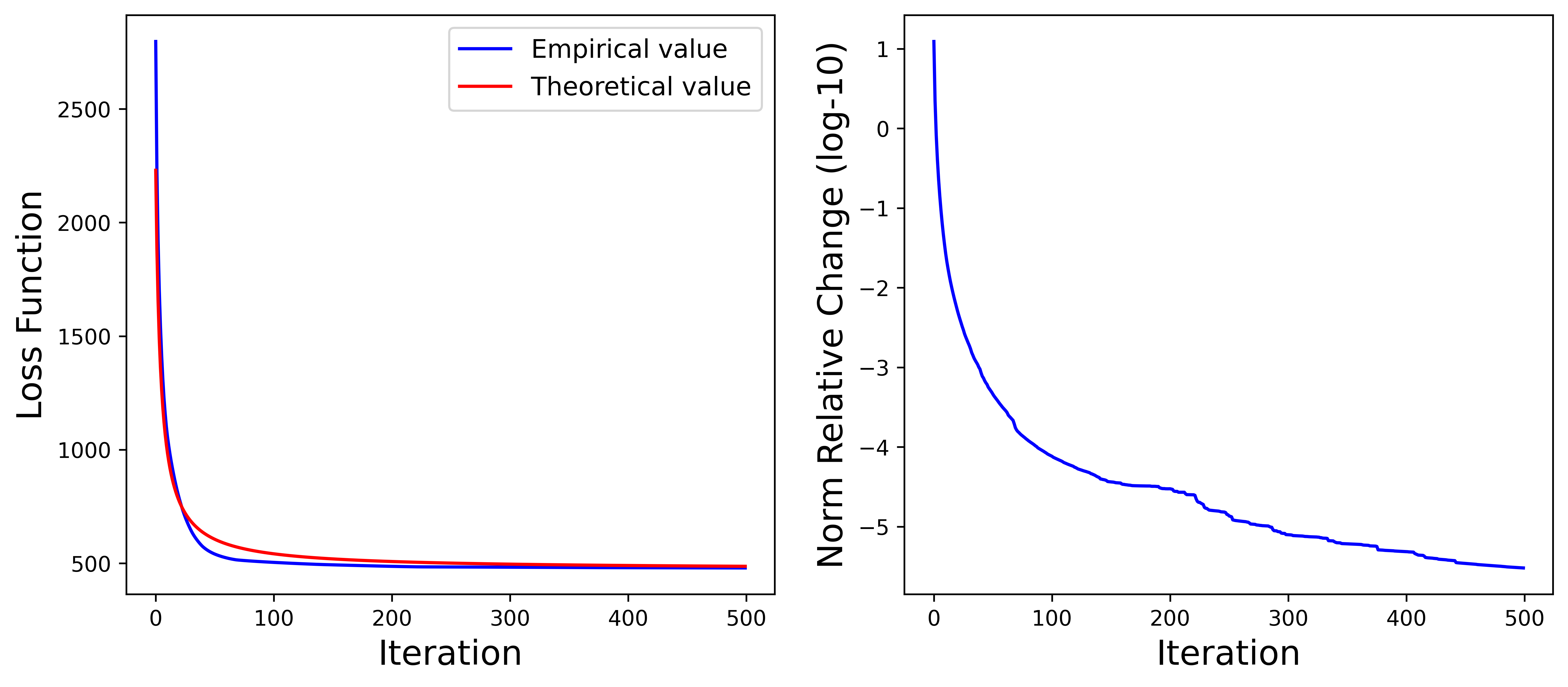}
    \caption{(Left) Loss history of the solar flare intensity regression task with Tensor-GPST ($\lambda = 1.0$); A curve at the order of $\mathcal{O}(1/K)$ is fitted to the loss history and empirically, the algorithm converges at the rate of $\mathcal{O}(1/K)$ to a local minimum. (Right) History of the Frobenius norm of the relative change of model parameters in log-10 scale, which suggests that the parameters converge to a stationary point, and thus the ALT-gap and the TV-gap will converge to a constant.}
    \label{fig:loss-function-example}
\end{figure}

Although we cannot fully verify the assumptions made, we plot the history of the loss function and the relative change of the model parameters for our real data application in Figure \ref{fig:loss-function-example}. Empirically, our model demonstrates a convergence rate at $\mathcal{O}(1/K)$ (see the red curve fitted based on a polynomial model with function form $f(k) = a + \frac{b}{c+k}$).

\section{Details of the Simulation Study}\label{app:simulation}
\begin{figure}[H]
    \centering
    \savebox{\largestimage}{\includegraphics[height=.22\textheight]{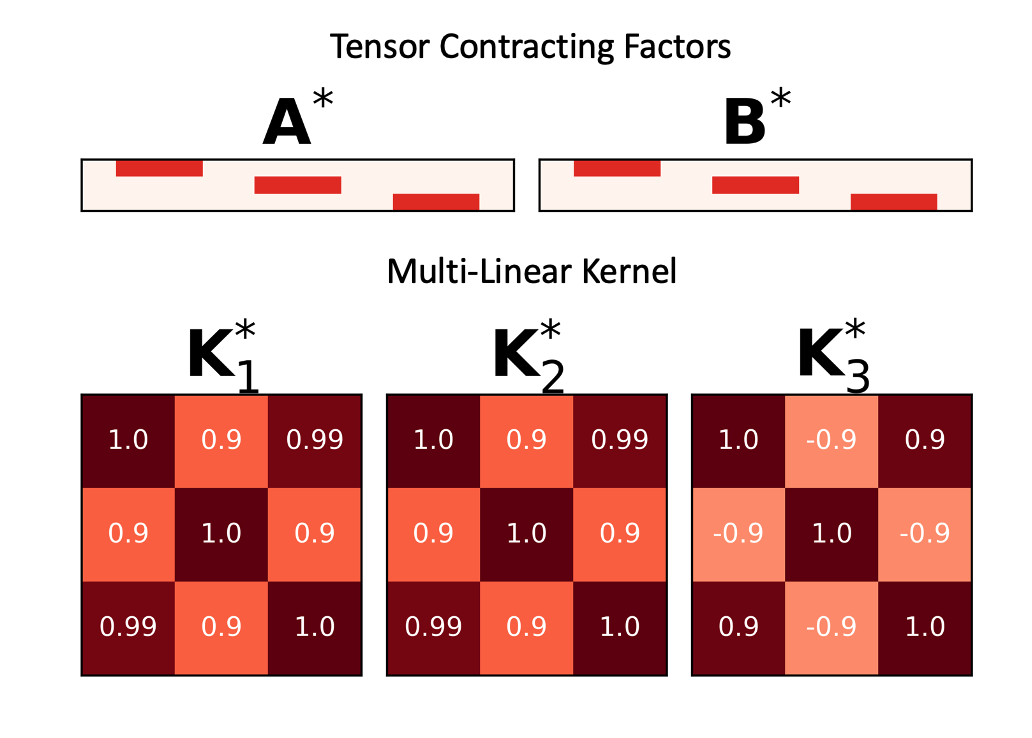}}%
    \begin{subfigure}[b]{0.48\textwidth}
        \centering
        \usebox{\largestimage}
        \caption{True tensor contracting factors (top) and true multi-linear kernels (bottom).}
        \label{fig:Simulation-AB-Kernel}
    \end{subfigure}
    \begin{subfigure}[b]{0.48\textwidth}
        \centering
        \raisebox{\dimexpr.5\ht\largestimage-.5\height}{%
      \includegraphics[width=0.98\textwidth]{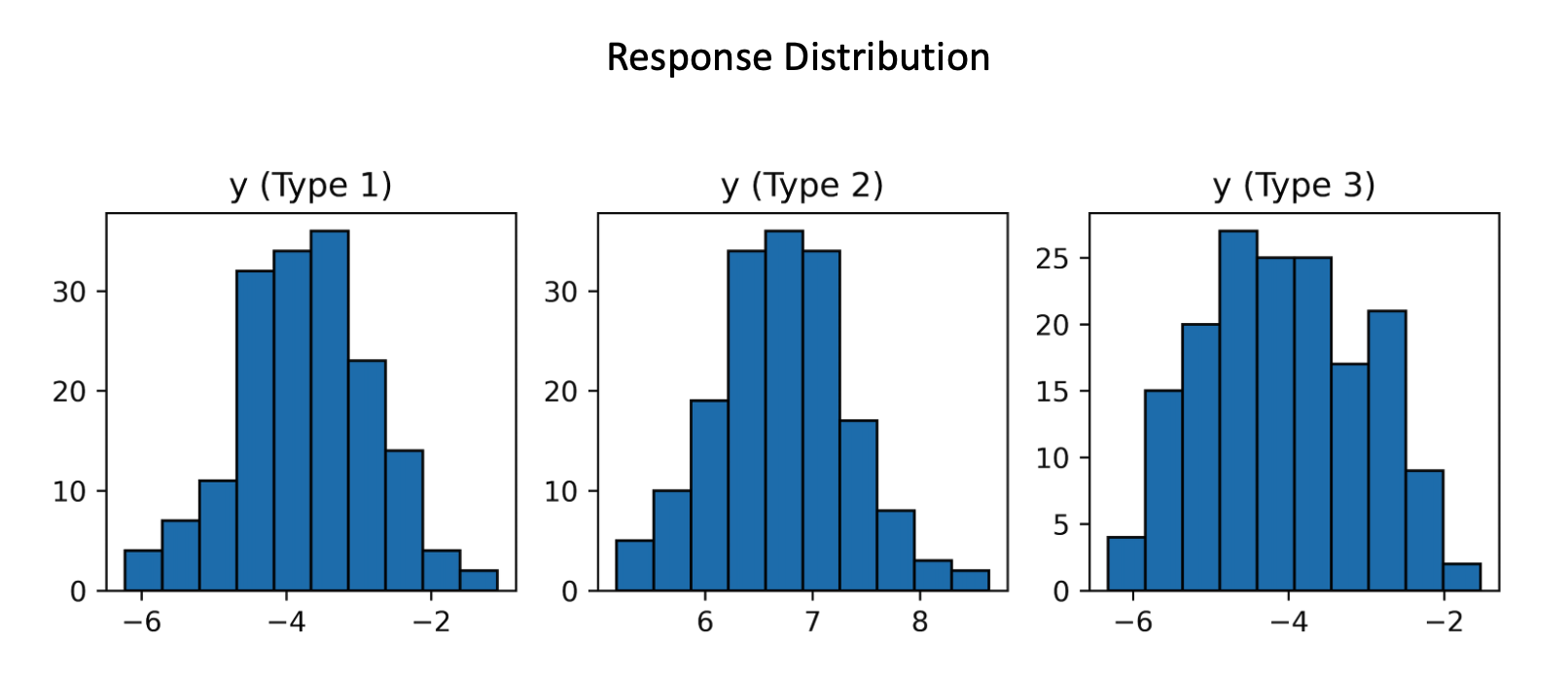}}
        \caption{Distribution of the response variable $y$ by type (see type definition in Figure \ref{fig:simdata}. Total sample size $N=500$.}
        \label{fig:Simulation-Response}
    \end{subfigure}
    \caption{Ground Truth of the Simulated data. (a) The true tensor contracting factors $(\mathbf{A}^{*},\mathbf{B}^{*})$ (top), where each has a banded structure with the 5 consecutive pixels filled with $0.2$ on each row. The bottom shows the multi-linear kernel $\mathbf{K}_{1}^{*},\mathbf{K}_{2}^{*},\mathbf{K}_{3}^{*}$. (b) The resulting response distribution of each type of data. One can see how type 1 \& 3 has similar distribution, thanks to their high channel correlation in $\mathbf{K}_{3}^{*}$.}
    \label{fig:Simulation-Truth}
\end{figure}

In this section we provide the details on generating the simulation data. Given the three types of data in Figure \ref{fig:simdata}, we use two sparse and banded tensor contracting factors $(\mathbf{A}^{*},\mathbf{B}^{*})$ (see the top of Figure \ref{fig:Simulation-AB-Kernel}) to contract each channel to a $3\times 3$ tensor (see Figure \ref{fig:tensor-contract} about the contraction operation). $(\mathbf{A}^{*},\mathbf{B}^{*})$ in Figure \ref{fig:Simulation-AB-Kernel} essentially do a $5\times 5$ block averaging for the four corners, four sides and the middle block of each channel data. So one can expect the Type 1 \& 3 data in Figure \ref{fig:simdata} to have its \textit{signal} in the four corners of the contracted tensor, and Type 2 has its \textit{signal} in the middle block (see Figure \ref{fig:simdata} for an illustration). Given the contracted tensor, we use the multi-linear kernel, specified in Figure \ref{fig:Simulation-AB-Kernel} (bottom) to generate the response variables via:
$$
\mathbf{y} \sim \mathcal{N}\left(\mathbf{0}_{N}, \mathbf{K}^{*}+\sigma^{2}\mathbf{I}_{N}\right)
$$
where $\mathbf{K}^{*}(i,j) = \vect{\mathcal{X}_{i}}^{\top}\left[\mathbf{K}_{3}^{*}\otimes \left(\mathbf{B}^{\top}\mathbf{K}_{2}^{*}\mathbf{B}\right) \otimes \left(\mathbf{A}^{\top}\mathbf{K}_{1}^{*}\mathbf{A}\right)\right]\vect{\mathcal{X}_{j}}$ and $\sigma=0.5$.

One can notice from the kernel $\mathbf{K}_{3}^{*}$ in Figure \ref{fig:Simulation-AB-Kernel} that channel $1$ \& $3$ are positively correlated and channel $2$ is negatively correlated with both channel $1$ \& $3$, and this is reflected in Figure \ref{fig:Simulation-Response}, where we plot the distribution of the simulated sample of size $N=500$, by the type of data. The tensor regression problem is to use the original $25\times 25\times 3$ tensors $\mathcal{X}_{i}$ to forecast the regression label $y_{i}$.

\section{Details of the AIA-HMI Solar Flare Imaging Dataset}\label{app:flare-dataset}
In this appendix, we provide some astrophysical backgrounds and additional details on data preprocessing about the AIA-HMI solar flare imaging datasets. 

There are over $12,000$ solar flares recorded by the Geostationary Operational Environmental Satellite (GOES) from May, $2010$ to June, $2017$, with intensity at least at the A-class flare level (peak X-ray brightness $< 10^{-7} \mbox{W/m}^{2}$). Among these flares, $4,409$ are B-class flares ($10^{-7} \sim 10^{-6} \mbox{W/m}^{2}$), $710$ are M-class flares ($10^{-5} \sim 10^{-4} \mbox{W/m}^{2}$) and $50$ are X-class flares ($> 10^{-4} \mbox{W/m}^{2}$). We combine the M-class and X-class flares in a single class, we name the class as M/X-class flares. Each flare is associated with a solar active region, which is a localized, transient volume of the solar atmosphere characterized by complex magnetic fields. We collect the AIA and HMI imaging data for each of the M/X-class flare during this period, and collect the B-class flares happened within the same active regions to construct our own database. Given the data availability, we end up with a database of $1,264$ B-class flares and $728$ M/X-class flares. 

\begin{figure}[htb]
    \centering
    \includegraphics[width=0.95\textwidth]{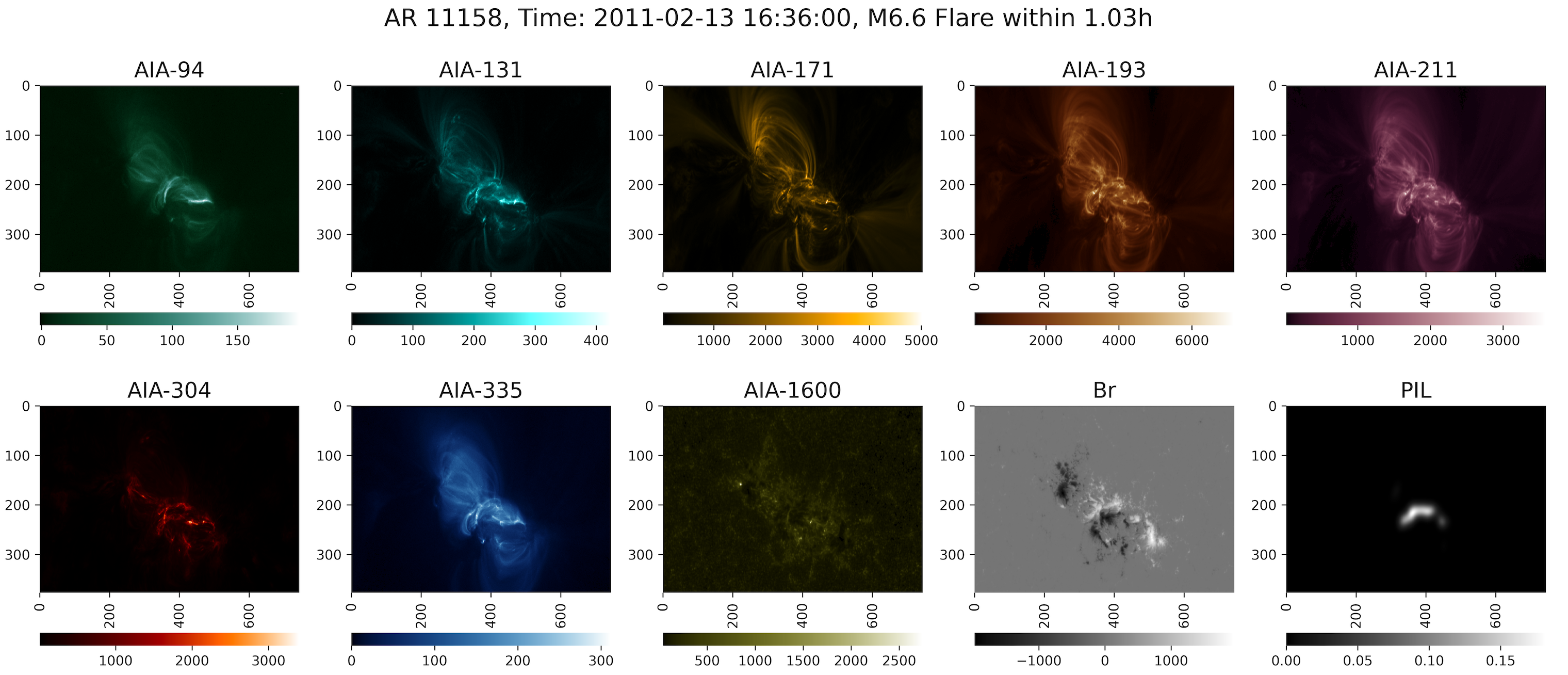}
    \caption{M-class Flare Example for Active Region (AR) No.11158, recorded at 16:36:00 (UT) of Feb 13, 2021. The flare intensity is $6.6\times 10^{-5} \mbox{W/m}^{2}$ and peaked at 17:38:00 (UT) of the same day. Tensor data size is $377\times 744 \times 10$. Channel name labeled on top of each panel where we omit the $\angstrom$.}
    \label{fig:flare-example-original}
\end{figure}

The AIA imaging data has 8 channels, distinguished by the wavelength band of the Extreme Ultraviolet (EUV) and Ultraviolet (UV) spectrum used to image the Sun \footnote{See more details at \url{https://sdo.gsfc.nasa.gov/data/channels.php}}. The AIA channels are named under their respective spectral band: AIA-$94 \angstrom$,AIA-$131 \angstrom$, AIA-$171 \angstrom$, AIA-$193 \angstrom$, AIA-$211 \angstrom$, AIA-$304 \angstrom$, AIA-$335 \angstrom$ and AIA-$1600 \angstrom$. The HMI imaging data captures the $r,\theta,\phi$-component of the solar magnetic field, and in our database, we keep the HMI  $\mbox{B}_{\mathrm{r}}$ channel, which have demonstrated contains flare-predictive signals \cite{sun2022predicting}. Finally, we derive the polarity inversion line (PIL) \cite{schrijver2007characteristic} from the $\mbox{B}_{\mathrm{r}}$, which highlights a sub-region with the strongest flare discriminating signals \cite{wang2020predicting,sun2021improved} and $\mbox{B}_{\mathrm{r}} \approx 0$. In Figure \ref{fig:flare-example-original}, we plot one example of the 10 channels for an M-class flare.

For the particular case in Figure \ref{fig:flare-example-original}, the image size is $377\times 744$, but different active regions are of different size. Also, different flares have their PIL, as well as the major signals in the other channels, stretching in different directions. To unify the size and orientation of all flares' imaging data, we follow these steps to preprocess our data:

\begin{itemize}
    \item Pick the pixel in the PIL channel with the largest sum of PIL weights near its $51\times 51$ neighborhood. This helps on picking the ``center" of the image. If the PIL only contains zeros, which could happen for some very weak B-class flares, we use the AIA-$1600\angstrom$ in place of the PIL and do the same thing.
    \item Around the ``center", we randomly sample $5,000$ pixels, with replacement and the sampling probability is proportional to the PIL (or AIA-$1600\angstrom$) pixel intensity, and do a Principal Component Analysis (PCA) of each pixel's 2D $(x,y)$ coordinates (coordinates on the pixel grid) and use the first principal component to calculate the orientation of the PIL. This step helps to find the ``direction" of the image.
    \item We rotate each channel with the same angle such that the ``direction" of the PIL is vertical. Then, we crop a $201\times 201$ window around the ``center" of the image, and do zero-padding where it is needed.
\end{itemize}

\begin{figure}[t!]
    \centering
    \includegraphics[width=0.95\textwidth]{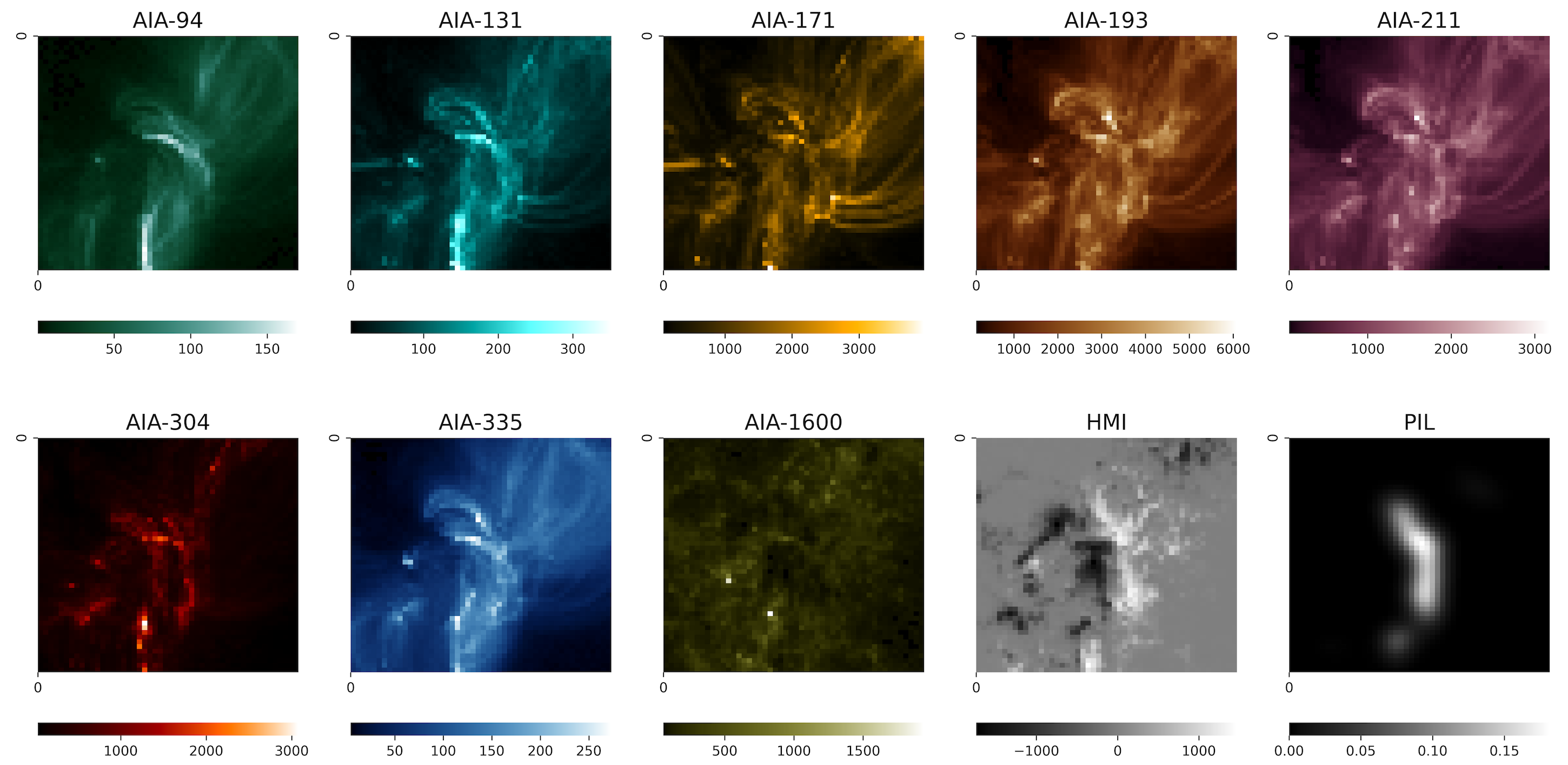}
    \caption{Pre-processed version of the sample in Figure \ref{fig:flare-example-original}. Notice how the PIL channel is now aligned vertically. Tensor size is reduced to $50\times 50\times 10$ for all $1,329$ flares.}
    \label{fig:flare-example-reduced}
\end{figure}

These preprocessing steps create flare data that are roughly comparable, but just as the simulation data pattern in Figure \ref{fig:simdata}, there is still randomness w.r.t. the positioning and direction of the flare predictive signal for each individual flare. We subset our flare list to those whose longitude is within $\pm 60^{\circ}$ from the Sun's central meridian, which removes the low-quality samples with limb distortion. This reduces our sample size from $1,992$ flares to $1,329$ flares. We further reduces the dimensionality of the $201\times 201$ images to $50\times 50$, after applying the preprocessing steps above, by bi-linear interpolation to speed up the model computation. The pre-processed version of the sample in Figure \ref{fig:flare-example-original}, with tensor size $50\times 50\times 10$, is shown in Figure \ref{fig:flare-example-reduced}. Notice how the PIL channel is now looking more ``vertical" and how each channel is sort of ``zoomed-in". The tensor size is now unified across all samples as $50\times 50\times 10$.

Before running the model, we normalize the scale of each channel such that each channel has its pixel intensity roughly within the range of $[-1,1]$, to avoid numerical overflow in the algorithm. We only use the training set scale information to determine the scaling factor in order to avoid information spillover.

For the flare intensity, originally, B-class flare has its intensity within the interval $[10^{-7},10^{-6}]$ (unit: $\mbox{W/m}^{2}$), and M/X-class flare has its intensity within the interval $[10^{-5},+\infty]$ (unit: $\mbox{W/m}^{2}$). We transform any intensity $y$ of each flare via:
$$
\tilde{y} = \log_{10}(y) + 5.5
$$
such that the middle point of the weakest M/X-class flare and the strongest B-class flare is centered at zero.

\section{Additional Results on Solar Flare Forecasting}\label{app:more-flare-results}
\begin{figure}[htb]
    \centering
    \includegraphics[width=0.98\textwidth]{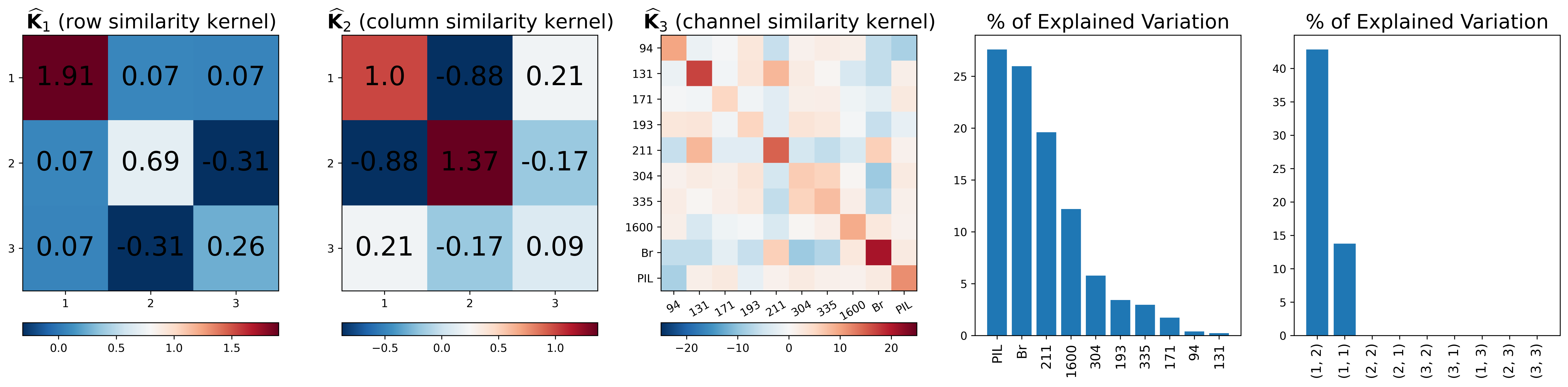}
    \caption{\textbf{GPST} (under random train/test split, $\lambda=1.0$) kernel estimates (panel 1-3), channel-wise $\%$ of explained variations (panel 4) and feature map $\%$ of explained variations (panel 5). It coincides with the literature \cite{wang2020predicting,sun2021improved} that the PIL is the channel with strong flare signals and the AIA imaging data is a good add-on to the HMI channel. The index for feature maps are the 2-tuple $(s,t)$.}
    \label{fig:kernel_est_GPST}
\end{figure}

\begin{figure}[htb]
    \centering
    \includegraphics[width=0.98\textwidth]{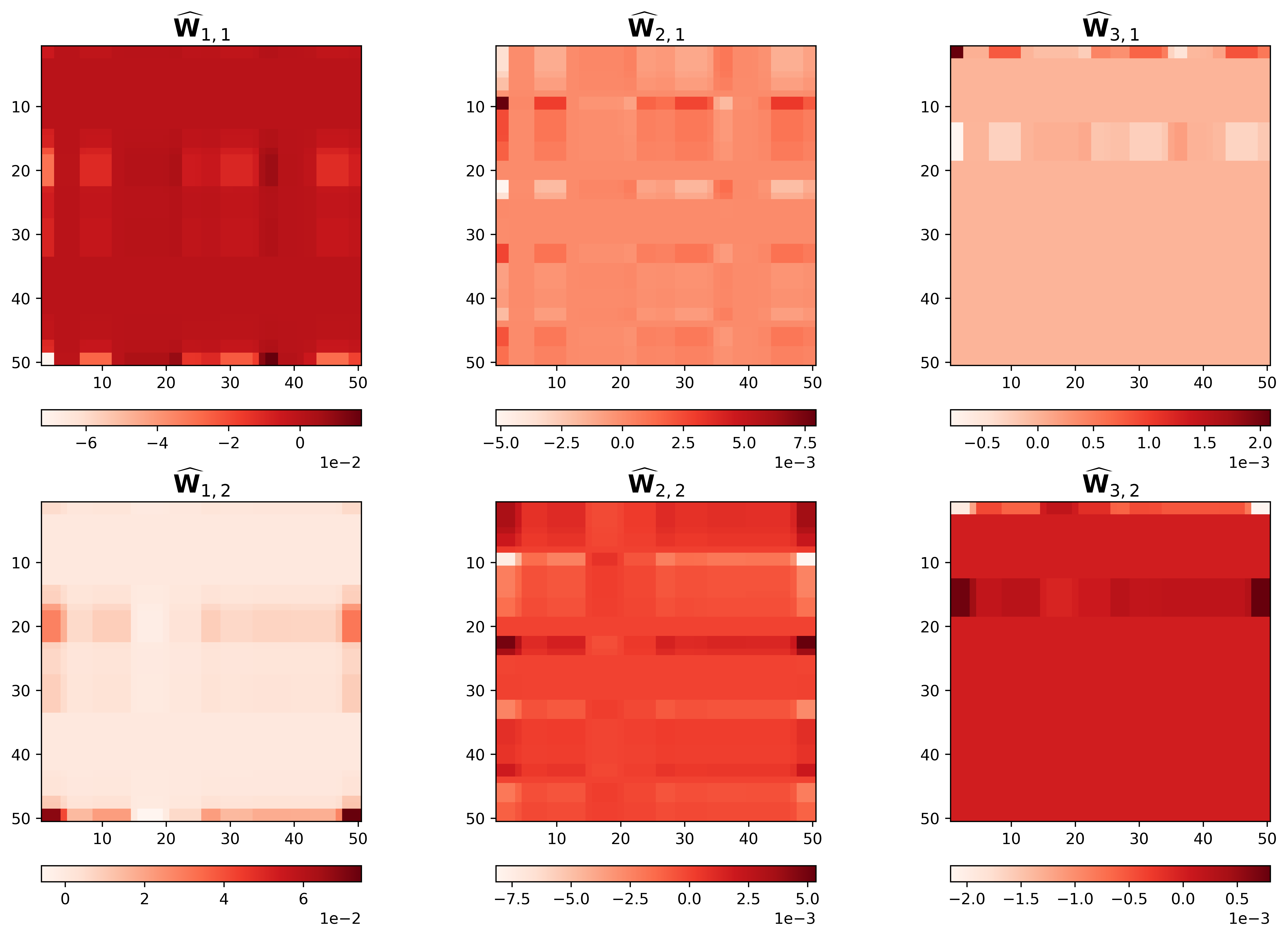}
    \caption{\textbf{GPST} (under random train/test split, $\lambda=1.0$) feature map (the non-zero ones) estimates.}
    \label{fig:feature_map_est_GPST}
\end{figure}

This appendix provides additional results on the solar flare intensity regression. We first visualize the parameter estimates of \textbf{GPST}, which is the best-performing model in Table \ref{tab:flare-results}, under one random train/test split with $\lambda=1.0$ in this section. Figure \ref{fig:kernel_est_GPST} provides the kernel estimates (the left three panels) and Figure \ref{fig:feature_map_est_GPST} shows the non-zero feature maps.

The kernel estimators $\widehat{\mathbf{K}}_{1}$ and $\widehat{\mathbf{K}}_{2}$ indicate that feature map $\mathbf{W}_{1,2}$ is of great importance since $\widehat{\mathbf{K}}_{1}(1,1)$ and $\widehat{\mathbf{K}}_{2}(2,2)$ contains the largest element, indicating that the feature extracted by $\mathbf{W}_{1,2}$ explains the most variations across all feature maps. 

To formally conceptualize the notion of feature map importance as well as channel importance, one can start by decomposing the variations of the regression label $y$ given tensor data $\mathcal{X}\in \mathbb{R}^{H\times W\times C}$ as follows:
\begin{equation}\label{eq:var-decomp}
\mbox{Var}(y) = \sum_{\substack{1\le s_1, s_2 \le h \\ 1\le t_1, t_2 \le w \\ 1 \le c_1, c_2 \le C}} \underbrace{\mathbf{K}_{1}(s_{1},s_{2})\cdot \mathbf{K}_{2}(t_{1},t_{2})}_{\text{\makebox[0pt]{Feature Map Importance}}}\times\overbrace{\mathbf{K}_{3}(c_1,c_2)}^{\text{\makebox[0pt]{Channel Importance}}}\times \underbrace{\left<\mathbf{W}_{s_1,t_1},\mathbf{X}^{(c_1)}\right>\cdot \left<\mathbf{W}_{s_2,t_2},\mathbf{X}^{(c_2)}\right>}_{\text{\makebox[0pt]{Latent Features Product}}} + \overbrace{\sigma^{2}}^{\text{\makebox[0pt]{Noise}}},
\end{equation}
and this leads to a natural definition of the percentage of explained variation for any \textit{pair} of channels $c_1, c_2 \in [C]$:
\begin{equation}\label{eq:single-channel-expvar}
{\%}\text{ Explained Variation} = \frac{\mathbf{K}_{3}(c_1,c_2)}{\mbox{Var}(y)}\times \sum_{\substack{1\le s_1, s_2 \le h \\ 1\le t_1, t_2 \le w}} \left<\mathbf{W}_{s_1,t_1},\mathbf{X}^{(c_1)}\right>\cdot \left<\mathbf{W}_{s_2,t_2},\mathbf{X}^{(c_2)}\right> \times \mathbf{K}_{1}(s_{1},s_{2})\cdot \mathbf{K}_{2}(t_{1},t_{2})
\end{equation}
Similarly, one can define the percentage of explained variation for any $pair$ of feature maps $\mathbf{W}_{s_1,t_1}$, $\mathbf{W}_{s_2,t_2}$, with $s_1, s_2 \in [h]$, $t_1, t_2 \in [w]$, as:
\begin{equation}\label{eq:single-map-expvar}
{\%}\mbox{ Explained Variation} = \frac{\mathbf{K}_{1}(s_1,s_2)\times \mathbf{K}_{2}(t_1,t_2)}{\mbox{Var}(y)}\times \sum_{1\le c_{1},c_{2}\le C}\left<\mathbf{W}_{s_1,t_1},\mathbf{X}^{(c_{1})}\right>\cdot \left<\mathbf{W}_{s_2,t_2},\mathbf{X}^{(c_{2})}\right> \times \mathbf{K}_{3}(c_{1},c_{2})
\end{equation}

The analysis here is a by-product of the Tensor-GPST model and is similar to the Joint and Individual Variation Explained (JIVE) \cite{lock2013joint} analysis. Both \eqref{eq:single-channel-expvar} and \eqref{eq:single-map-expvar} can be computed empirically by plugging in the parameter estimators of $\widehat{\mathbf{K}}_{1},\widehat{\mathbf{K}}_{2},\widehat{\mathbf{K}}_{3}$ and $\widehat{\mathbf{W}}_{s,t}, s\in [h], t\in [w]$ and use all training inputs $\mathcal{X}$ for calculation and take an average.

In the last two panels of Figure \ref{fig:kernel_est_GPST}, we show the percentage of explained variation for all $10$ AIA-HMI channels based on \eqref{eq:single-channel-expvar} and the percentage of explained variation for all $9$ feature maps based on \eqref{eq:single-map-expvar}. For channel-wise explained variation, we simply fix $c_{1} = c_{2}$ in \eqref{eq:single-channel-expvar}, and for feature map explained variation, we simply fix $(s_{1},t_{1}) = (s_{2},t_{2})$. Note that since all channels share the same set of feature maps, the latent features of different channel are not orthogonal, which indicates that the sum of the percentage of explained variation defined in \eqref{eq:single-channel-expvar} could exceed $100\%$. The same argument holds for the feature maps' explained variation. But the explained variation still reveals the relative importance of channels and feature maps.

\begin{figure}[htb]
    \centering
    \includegraphics[width=0.98\textwidth]{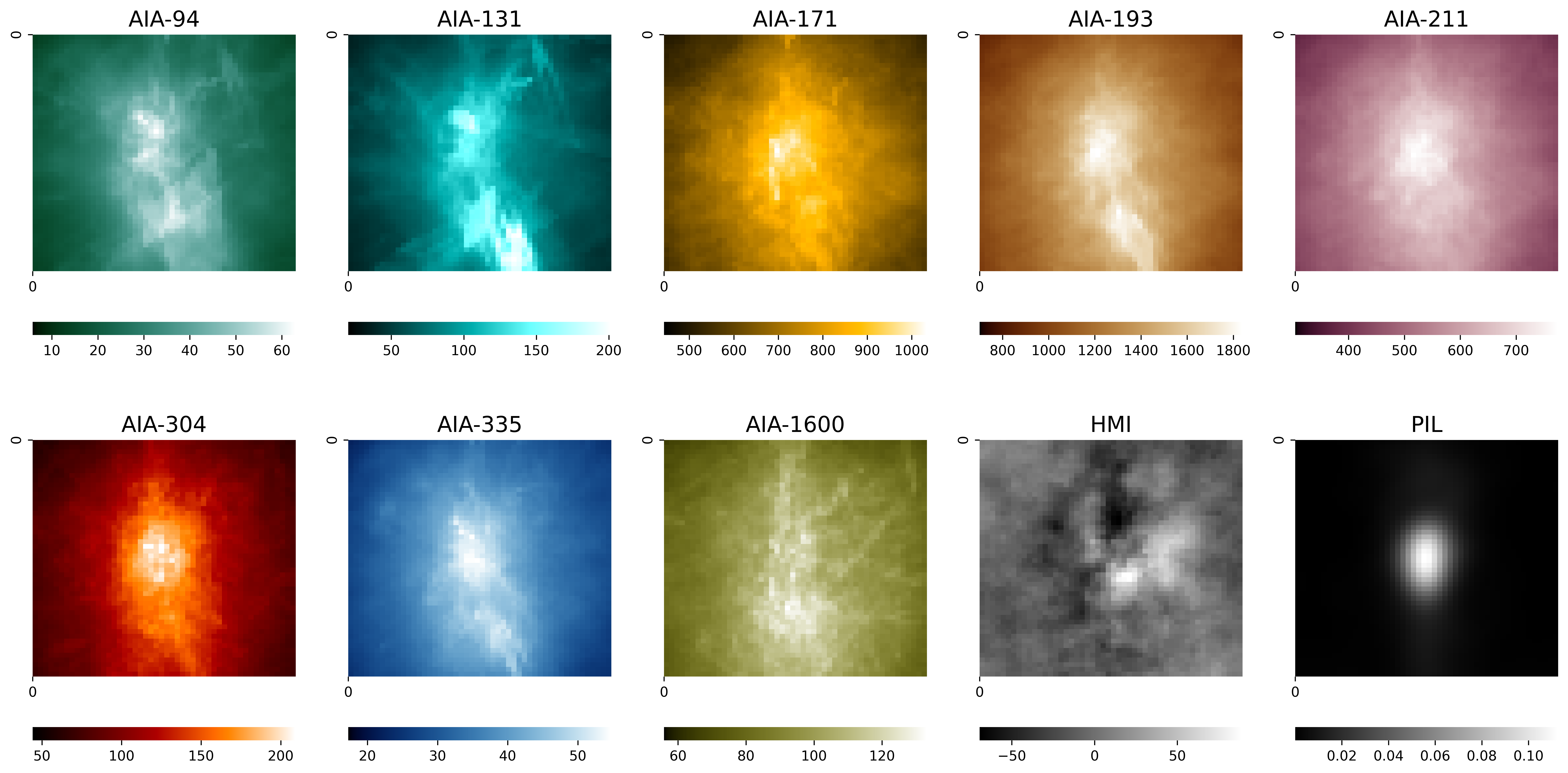}
    \caption{Sample average AIA-HMI map for M-class flare.}
    \label{fig:average-M-flare}
\end{figure}

\begin{figure}[htb]
    \centering
    \includegraphics[width=0.98\textwidth]{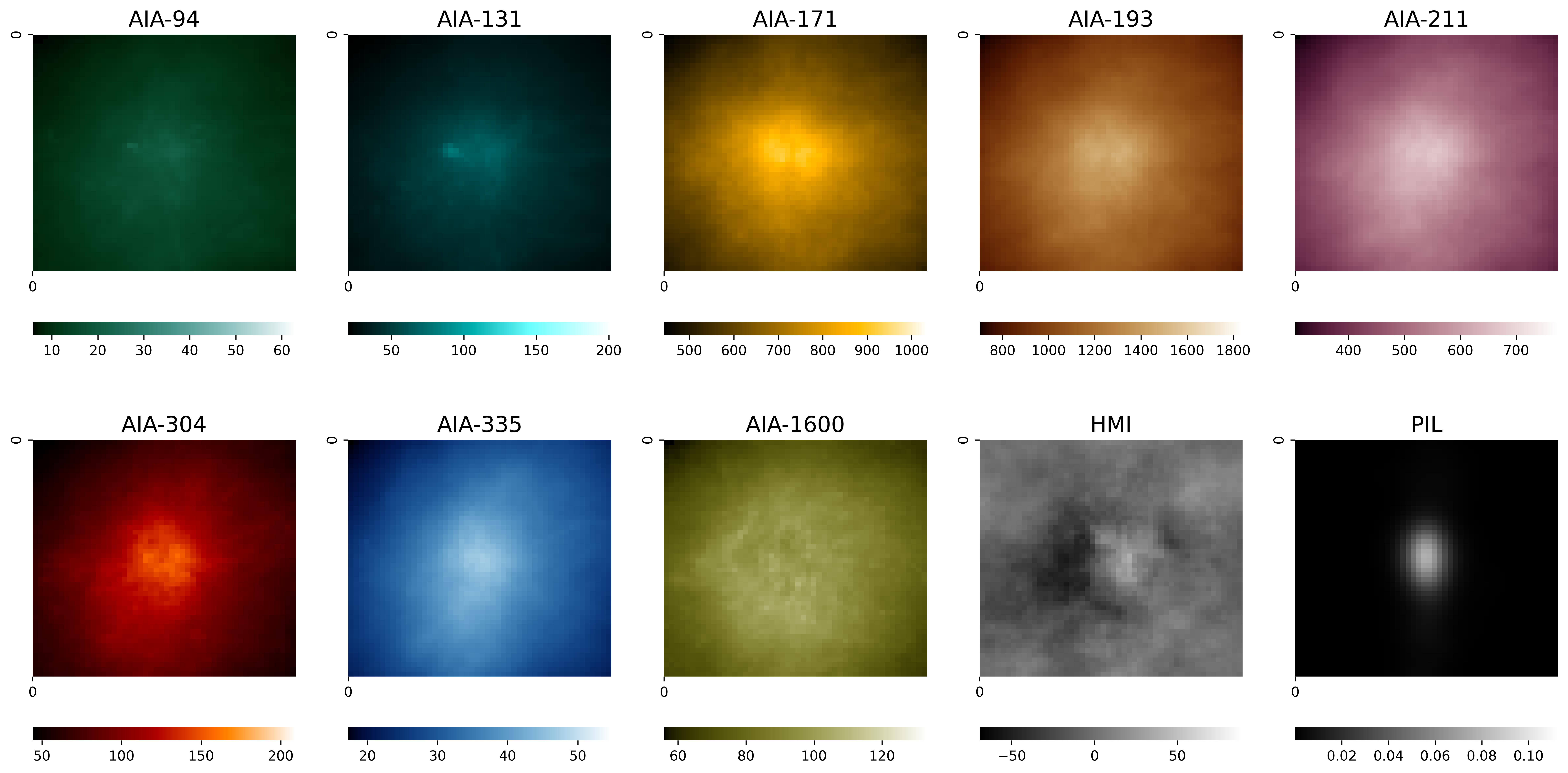}
    \caption{Sample average AIA-HMI map for B-class flare.}
    \label{fig:average-B-flare}
\end{figure}

The feature maps shown in Figure \ref{fig:feature_map_est_GPST} mainly highlight two patterns: 
\begin{itemize}
    \item All six feature maps show non-zero weights on at least one of the four boundaries. This indicates that the features collected are around the perimeter of the flare eruptive region, which captures the ``size" of the flare eruptive area. In Figure \ref{fig:average-M-flare} and \ref{fig:average-B-flare}, we show the sample average of all 10 channels for the M/X-class and B-class flares, respectively. One can easily notice the difference between the two classes in terms of the ``size" of the bright spots.
    \item There are some non-zero weights in $\mathbf{W}_{1,2}$ and other feature maps near row $20$, where features are collected near the top of the brightest PIL region of the M/X flares.
\end{itemize}

\end{document}